\documentclass[twocolumn]{aa} 
\usepackage{graphicx}
\usepackage{txfonts}
\usepackage{color}

\begin{document}
   \title{Dissecting the morphological and spectroscopic properties of galaxies in the local Universe}

   \subtitle{I. Elliptical galaxies}

   \author{J. A. L. Aguerri\inst{1,2}, M. Huertas-Company\inst{3,4}, J. S\'anchez Almeida\inst{1,2}  \and C. Mu\~noz-Tu\~n\'on\inst{1,2}
          }

   \institute{Instituto de Astrof\'{\i}sica de Canarias, C/ V\'{\i}a L\'actea s/n, 38200 La Laguna
    \and 
    Departamento de Astrof\'{\i}sica de la Universidad de La Laguna.
    \and
    GEPI, Paris-Meudon Observatory, 5 Place Jules Janssen, 92190 Meudon, France 
	\and
    Universit\'e Paris Diderot, 75205 Paris Cedex 13, France \\
     \email{jalfonso@iac.es; marc.huertas@obspm.fr; jos@iac.es; cmt@iac.es}
     }

   \date{Received XXX; accepted XXX}

 
  \abstract
  {}   
   {We revisit the scaling relations and star-forming histories of local elliptical galaxies using a novel selection method applied to  the Sloan Digital Sky Survey DR7. }
   {We combine two probability-based automated spectroscopic and morphological classifications of $\sim 600000$ galaxies with $z<0.25$ to isolate \emph{true} elliptical galaxies. Our sample selection method does not introduce artificial cuts in the parameters describing the galaxy but instead it associates to every object a weight measuring the probability of being in a given spectro-morphological class. Thus the sample minimizes the selection biases. }
   {We show that morphologically defined ellipticals are basically distributed in 3 spectral classes, which dominate at different stellar masses. The bulk of the population ($\sim 50\%$) is formed  by a well defined class of galaxies with old stellar populations that formed their stars at very early epochs in a short episode of star formation. They dominate the scaling relations of elliptical galaxies known from previous works and represent the \emph{canonical} elliptical class. At the low mass end, we find a population of slightly larger ellipticals, with smaller velocity dispersions at fixed stellar mass, which seem to have experienced a more recent episode of star formation probably triggered by gas-rich minor mergers.  The high mass end tends to be dominated by a third spectral class, slightly more metal rich and with more efficient stellar formation than the reference class. This third class contributes to the curvature of the mass-size relation at high masses reported in previous works. Our method is therefore able to isolate typical spectra of elliptical galaxies following different evolutive pathways.}
   {}

\authorrunning{J. A. L. Aguerri et al.}
\titlerunning{Properties of galaxies in the local Universe: I. Elliptical galaxies}
   \maketitle
%

\section{Introduction}

\indent

More than 50\% of the stellar mass in today's universe is in early-type galaxies (see \cite{fukugita1998}).  Understanding their formation and evolution is therefore one of the key questions in modern structure formation theories.

Basically, there is an open debate on whether actual elliptical galaxies are the end products of a long history of major and/or minor dry/wet galaxy mergers (see e.g. \cite{delucia2006}, hierarchical scenario) or if they are passively evolving galaxies formed at high redshift in a short time-scale with almost no accretion  of external material since their early formation (monolithic scenario; see e.g. \cite{kodama1998}, \cite{vandokum2003}, \cite{carretero2007}).  From the observational point of view, the scaling relations are a powerful tool to disentangle between these two evolutionary paths.

The passively evolving scenario or monolithic collapse model  (\cite{partridge1967}, \cite{larson1975}) is supported by the fact that early-type galaxies form, at first sight, an homogeneous family of objects with very regular morphology (they do not show strong spiral arms or asymmetries like other types of galaxies; e.g. Abraham et al. 1996). In addition, their surface brightness distribution is concentrated towards the center, and can be well fitted by a Sersic surface brightness profile (\cite{caon1993, trujillo2001, graham2003, aguerri2004, gutierrez2004, trujillo2004, kormendy2009}). This regular morphology is also reflected in the color (i.e., stellar populations). Early-type galaxies form indeed a well defined family of galaxies in the color-magnitude diagram on the so-called red sequence (\cite{kodama1997,bell2004,balogh2004}) which suggests that they are formed by old stellar populations of similar age with almost no star formation activity at present. This holds at least up to redshift 1 (\cite{mei09}) and even up to redshift 2 (\cite{andreon2011}). From the dynamical point of view, they are dominated by random motions with a high degree of virialization (see e.g. \cite{kronawitter2000}, \cite{gerhard2001}, \cite{thomas2007}). They consequently follow tight scaling relations defined by their size, velocity dispersion, surface brightness, and luminosity. The most commonly studied relations are: the luminosity/mass-size (\cite{shen2003}, \cite{mcintosh2005}), the Faber-Jackson relation (\cite{faber1976}), the Kormendy relation (\cite{kormendy1977}), and the fundamental plane (\cite{djorgovsky1987}, \cite{dressler1987}). 

Detailed studies on scaling relations of early-type galaxies have shown that despite of the apparent homogeneity, there are important differences among them. Some works have in fact found that the scaling relations and/or structural parameters of early-type galaxies depend on different galaxy properties, such as: luminosity (\cite{bender1992}, \cite{caon1993},  \cite{desroches2007}, \cite{nigoche2008}, \cite{aguerri2009},\cite{nigoche2009}, \cite{nigoche2010},\cite{bernardi2010}), stellar populations (\cite{forbes1998}, \cite{terlevich2002},\cite{graves2009}, \cite{graves2010}), or environment (\cite{aguerri2004}, \cite{gutierrez2004}, \cite{fritz2005}). Moreover, a strong size evolution is observed in early-type galaxies from high to low redshift (\cite{daddi2005, trujillo2006}). These differences can be only understood if galaxies with different luminosities, stellar populations or located in different environments have experienced different formation histories. According to these properties, some authors have seen a dichotomy in the early-type population, and divided them in two different families. One of the families  would be formed by the low luminous early-type galaxies. They show fast rotation, are isotropic oblate spheroids with discy isophotes, and show flat surface brightness profiles in their central regions (they are core-less objects). These galaxies are similar to bulges of spiral galaxies. In contrast, the more luminous or giant early-type galaxies are non-rotating, anisotropic and triaxial systems. In addition, they show cuspy cores and boxy isophotes. These galaxies are also older and more $\alpha$-enhanced than the less luminous counterparts (\cite{kormendy1982},  \cite{nieto1991}, \cite{bender1992}, \cite{kormendy1996}, \cite{faber1997}, \cite{lauer2005}, \cite{lauer2007}, \cite{emsellen2007}, \cite{cappellari2007}, \cite{kormendy2009}). This dichotomy has been challenged by other authors (\cite{jerjen1997}, \cite{graham2003}, \cite{gavazzi2005}). 

The differences reported above in early-type galaxies could be explained if their formation history is somehow more complex than what is suggested by the monolithic scenario. Since the pioneering work by Toomre \& Toomre (1972), it is thought that early-type galaxies can be formed by mergers of disk galaxies. Detailed simulations report that two equal disks do not survive after a merger process (\cite{barnes1991,hernquist1993,barnes1996,mihos1996,dimatteo2005,naab2006b,robertson2006,hopkins2009}). The end product of such mergers is a galaxy with early-type morphology. These numerical simulations have also shown that the differences observed in the properties of early-type galaxies can be explained through the different physical processes taking place in such kind of mergers. This involves gas dissipation, star formation, and super-massive black hole feedback (\cite{barnes1992}, \cite{barnes1996}, \cite{mihos1996}, \cite{dimatteo2005}, \cite{robertson2006}, \cite{naab2006a}, \cite{naab2006b}, \cite{naab2007}, \cite{hopkins2008}). In particular, the gas fraction in the galaxy progenitors could play a crucial role in shaping the final merger remnants and fix their scaling relations (see, e.g., \cite{robertson2006}, \cite{hopkins2008}). 

In addition, selection effects could be responsible of increasing the scatter of the observed scaling relations of early-type galaxies. Early-type galaxies are usually selected by their photometric properties, being the final samples contaminated by galaxies with discs (e.g. S0s) which are not pure spheroidal systems (see \cite{bernardi2010}). Color is also one of the most used selection criterion. However, it has been recently observed that the red sequence containts a large fraction ($\approx 50\%$) of red spiral galaxies (see, e.g., \cite{emsellen2007}, \cite{krajnovic2008}, \cite{mahajan2009}, \cite{masters2010}, S\'anchez-Almeida et al. 2011). The above contamination may explain differences in the reported scaling relations, since discs and spheroids follow different trends (see \cite{shen2003}).  

In the present paper, we revisit the scaling relations of red-sequence elliptical galaxies using a new approach applied to one of the largest dataset of galaxies available today: the spectroscopic sample of  the Sloan Digital Sky Survey Data Release 7 (SDSS-DR7). The new approach of the present work is probabilistic, based on the spectroscopic and morphological classifications recently developed by our group (\cite{sanchezalmeida2010}, \cite{huertas2011}). This double classification provides a robust sample of early-type galaxies avoiding the contamination by other morphological types. Thus, early-type galaxies are formed by three different classes of objects according to their spectra, and we study the scaling relations of these three populations. The differences observed are interpreted in terms of different formation/evolution mechanisms.

The paper is organized as follows. We describe the sample selection in section~\ref{sec:sample}. Scaling relations such as color-mass, size-mass, Faber-Jackson, and the fundamental plane are shown in Section 3. The discussion and conclusions are given in Sections 4 and 5, respectively. The cosmological parameters adopted through this paper were $\Omega_{m}=0.3$, $\Omega_{\Lambda}=0.7$, and $H_{0}=70$ km s$^{-1}$ Mpc$^{-1}$. This paper is focused on elliptical galaxies (excluding S0s). 


\section{Sample selection}
\label{sec:sample}
\indent

Our starting point for the sample selection is all galaxies in SDSS DR7 (\cite{abazajian2009}) with spectra of good quality (not too close to the edges of the survey, not saturated, and properly deblended), apparent magnitude in $r$ brighter than 17.8, and redshift below 0.25 (see \cite{sanchezalmeida2010} for a detailed description of the sample selection). As result, the final number of galaxies used in the present work is 698420.  

Traditionally, early-type galaxies have been visually selected. This can be easily done for small and nearby galaxy samples. Nevertheless, visual classifications of samples as large as the one we use are very time consuming, and only by involving a large number of classifiers  it can be done in a reasonable amount of time (e.g. Galaxy Zoo project \cite{lintott2008, lintott2011}). Automated galaxy classification techniques have been developed during the last decades in order to classify large databases. The baseline of these techniques is that relative simple criteria using global galaxy properties can separate different galaxy types. Thus, early-type galaxies have been selected according to: measurements of the surface brightness profiles (\cite{Aguerri2002},\cite{aguerri2004}, \cite{trujillo2004}, \cite{gutierrez2004}, \cite{mendez2008}), colors (\cite{strateva2001}), light concentration and asymmetry (\cite{abraham1996}), or some combination of photometric and spectroscopic information (\cite{bernardi2003}, \cite{nigoche2010}).

These classifications assume that the different types of galaxies are well separated in the phase space defined by the considered parameters.  Nevertheless, this is not the case. There is a continuous transition between the different galaxy types. This makes it difficult to define a pure elliptical galaxy sample since contaminations from other galaxy types are always present. In our present work, the selection of  elliptical galaxies has been done implicitly using a probability based approach not explored before. We use the full sample weighting the derived relationships with the probability of being elliptical, both morphologically and spectroscopically. Thus even if spirals and ellipticals with uncertain classification are included, they do not contribute to the results.

 The morphological classification is based on support vector machines (SVM). This is a machine learning algorithm that tries to find the optimal boundary between several clouds of points in an N-dimensional space (for more details, see \cite{huertas2008}).  The galaxies of the SDSS DR7 spectroscopic sample were classified in four different morphological types (E, S0, Sab, and Scd). We associated to each galaxy of the sample a probability of being in these four morphological classes (see \cite{huertas2011}). We have also classified the galaxy spectra using the k-means clustering algorithm. This algorithm assumes that each galaxy spectra is a vector in a multi-dimensional space, with as many dimensions as the number of spectral wavelengths. It is assumed that the vectors are clustered around a number of cluster centers. The k-means algorithm finds the number of clusters, the cluster centers, and assigns to each galaxy one of the clusters. The galaxies of the sample were finally classified in 17 major ASK spectral classes, and each galaxy has also a probability of belonging to all classes (see \cite{sanchezalmeida2010}). 

We have combined the available spectroscopic and morphological information by assigning a weight ($P_{i}(\rm{E})$) to each galaxy. This weight is the product of the probability of the i-th galaxy to be classified morphologically as elliptical ($P_{i,\rm{morph}}(\rm{E})$) and the probability of the galaxy to belong to the j-th spectral class ($P_{i,\rm{ASK}}(A_{j})$). Note that $P_{i}(\rm{E})$ would be a probability in the case that the two probabilities were independent. Otherwise, $P_{i}(\rm{E})$ is just a weight which combines the spectroscopic and morphological information for each galaxy. In the following, we will consider that, for each galaxy, the larger the value of $P_{i}(\rm{E})$, the higher the probability of being elliptical.

Early-type galaxy samples usually include both elliptical and S0 galaxies. Nevertheless, these two kinds of galaxies are dynamically different. In particular, S0 galaxies present a disc structure absent in ellipticals. This different dynamics can affect to the scaling relations of the galaxies. Then, differences observed in the scaling relations of early-type galaxies could be related to different dynamics, rather than due to different origins and/or evolutions (see e.g. \cite{persic1996}; \cite{bell2001}; \cite{shen2003}; \cite{courteau2007}). In order to have a sample of pure spheroidal systems, we have not included in $P_{i}(\rm{E})$ the probability of each galaxy to be classified as S0.

\begin{figure*}
\includegraphics[width=0.24\textwidth]{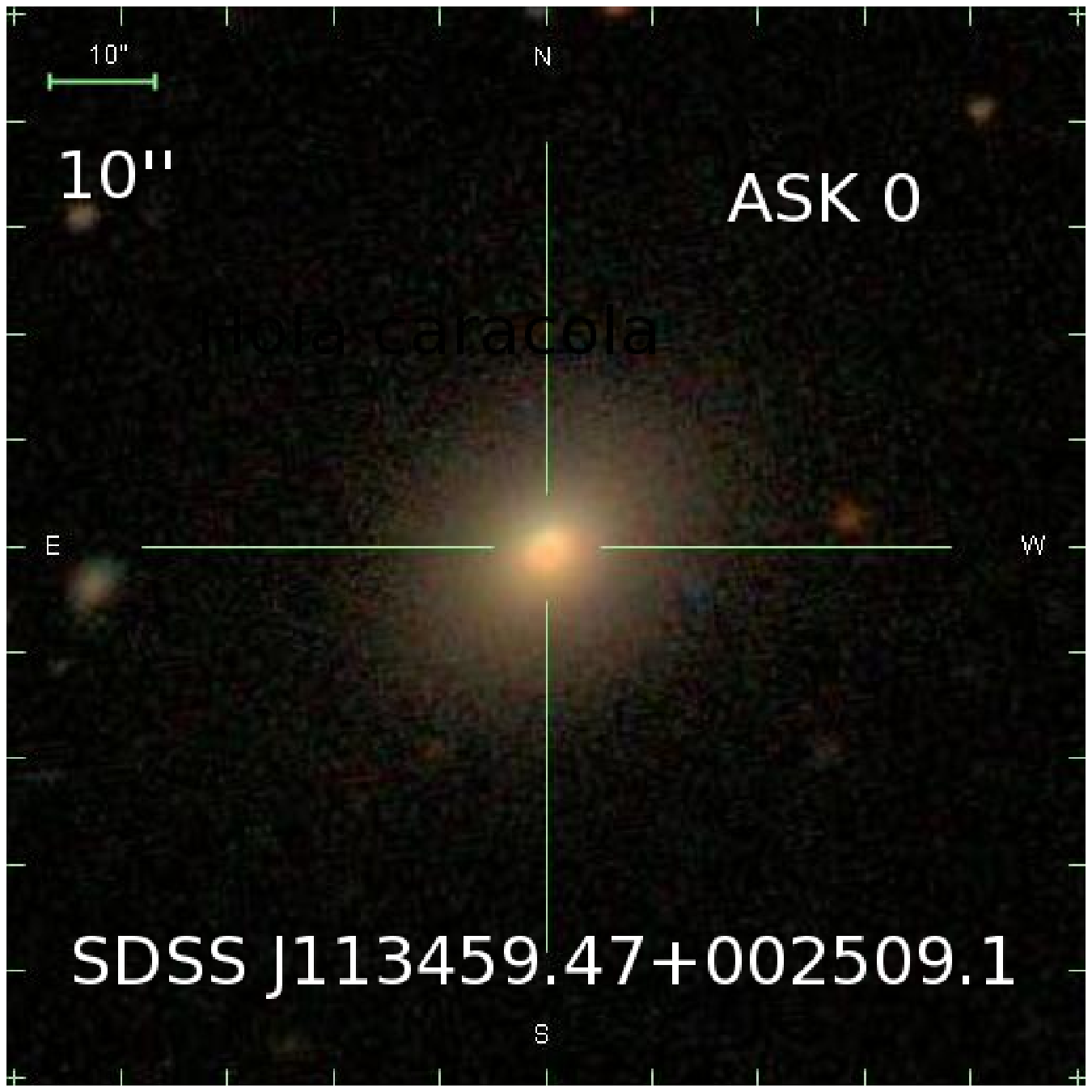}
\includegraphics[width=0.24\textwidth]{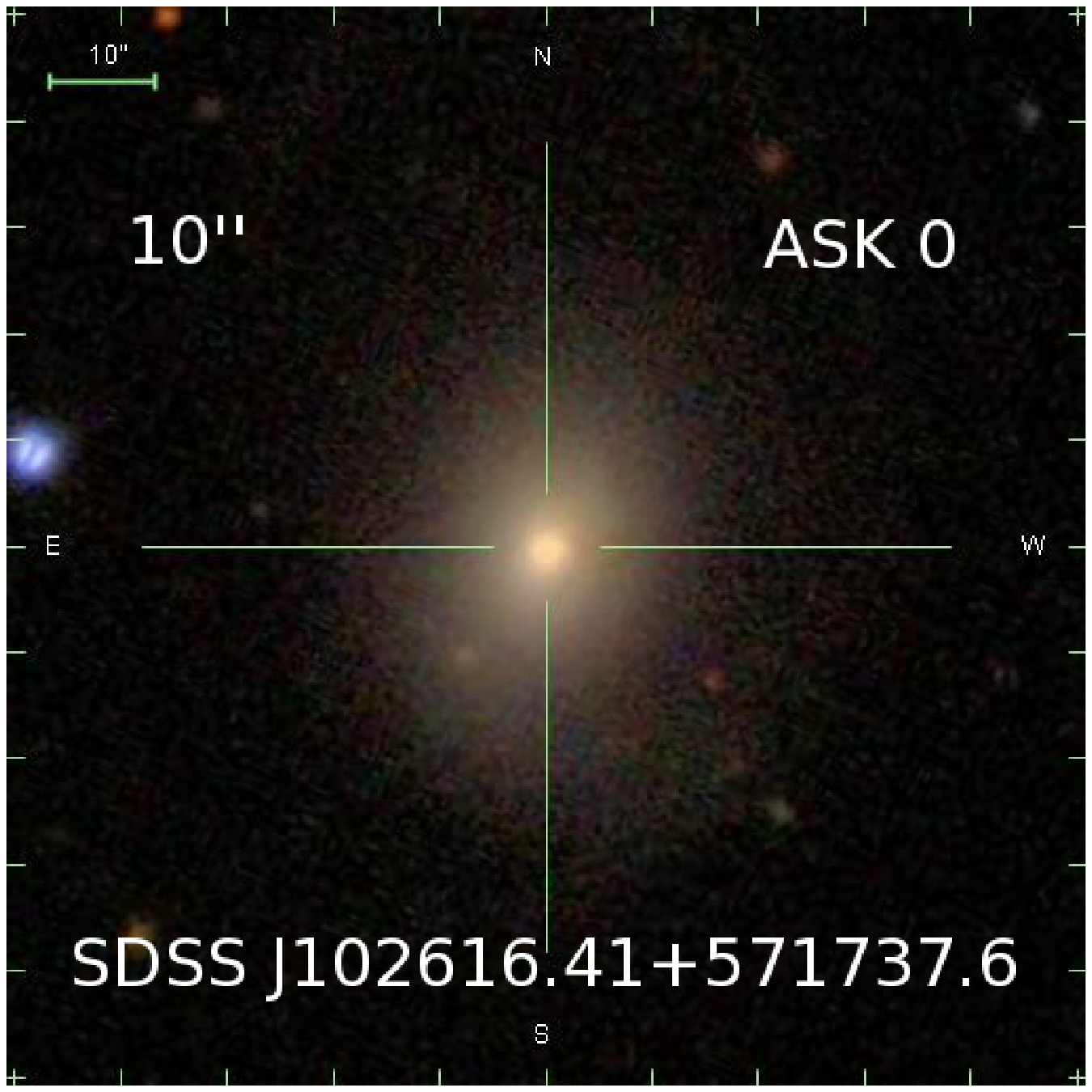}
\includegraphics[width=0.24\textwidth]{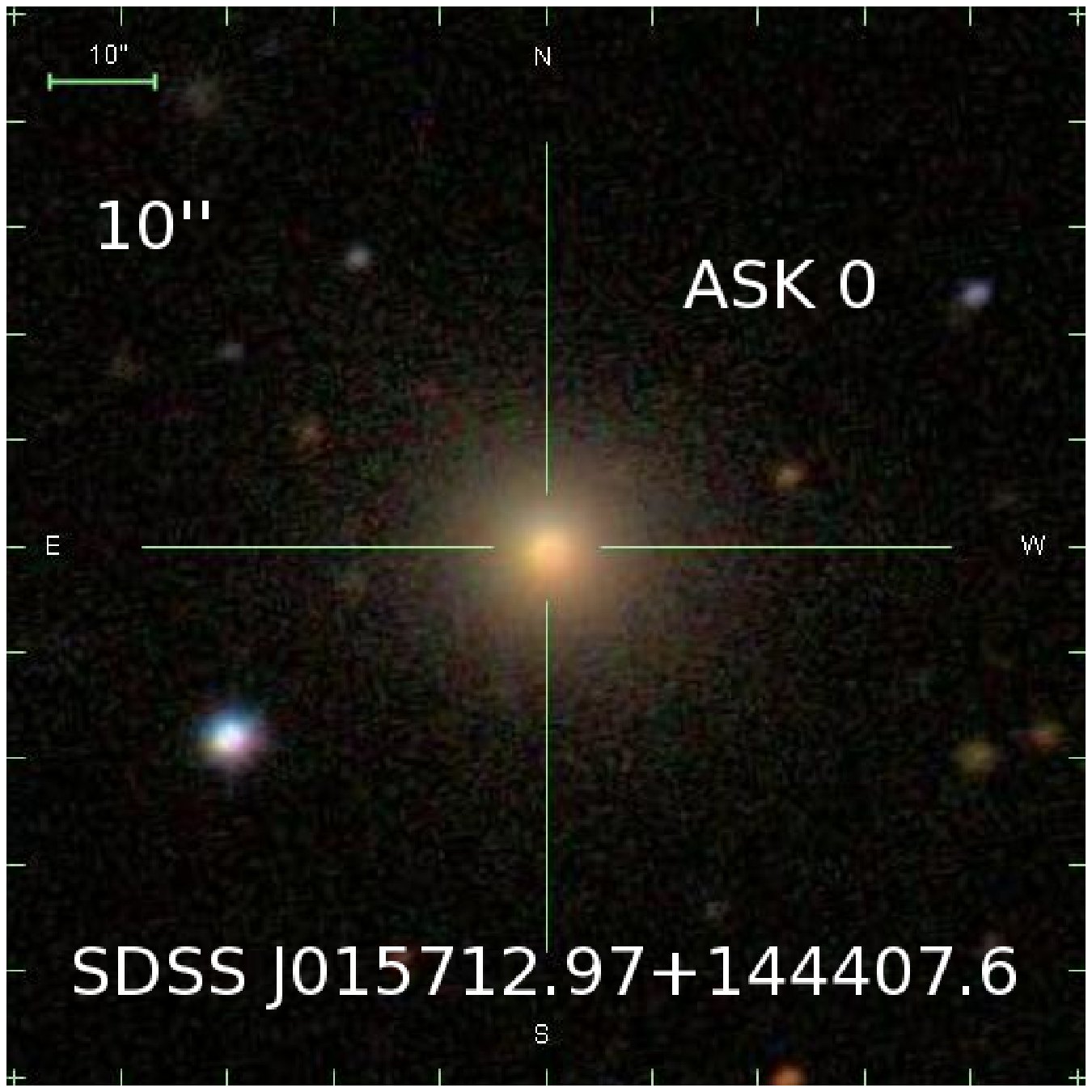}
\includegraphics[width=0.24\textwidth]{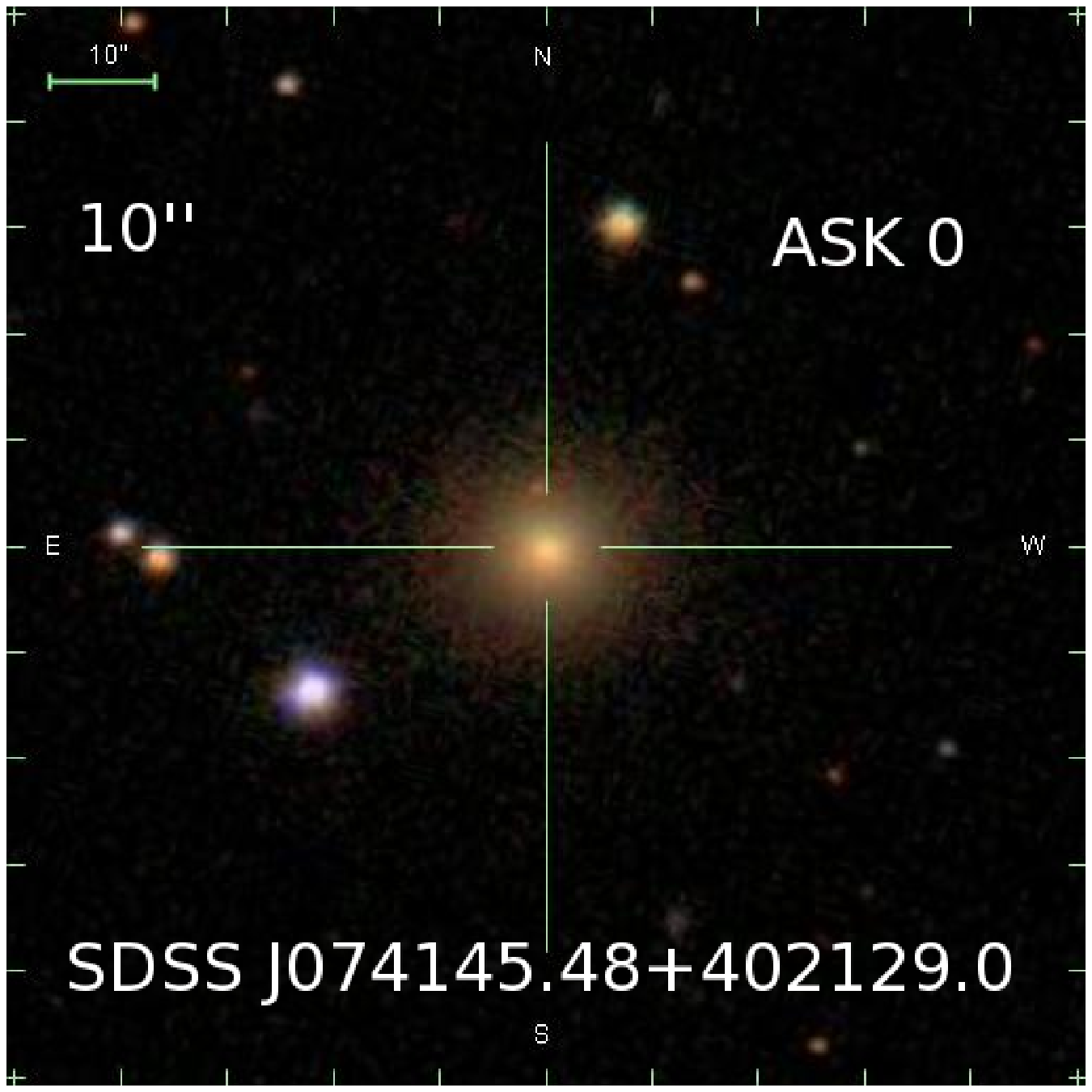}\\
\includegraphics[width=0.24\textwidth]{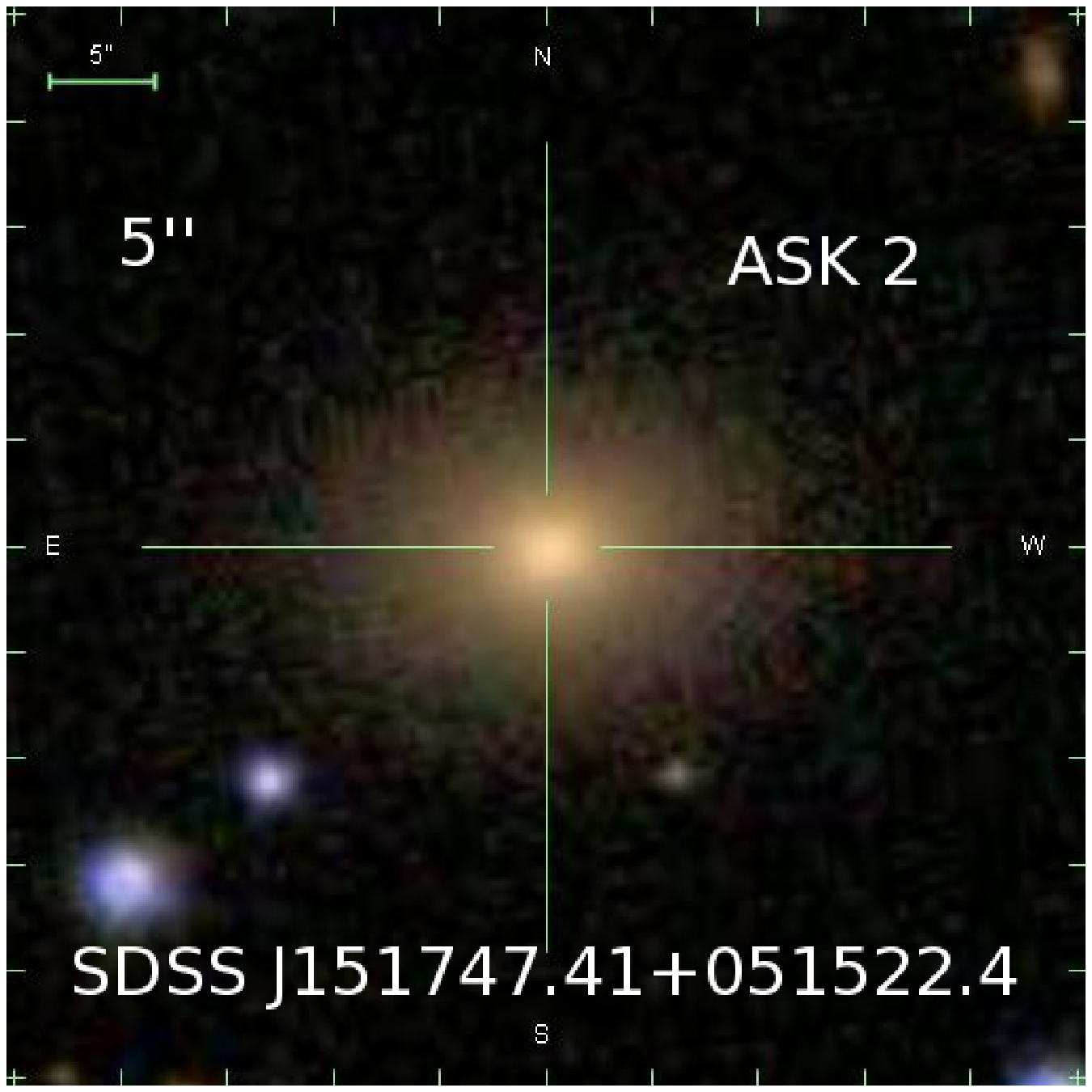}
\includegraphics[width=0.24\textwidth]{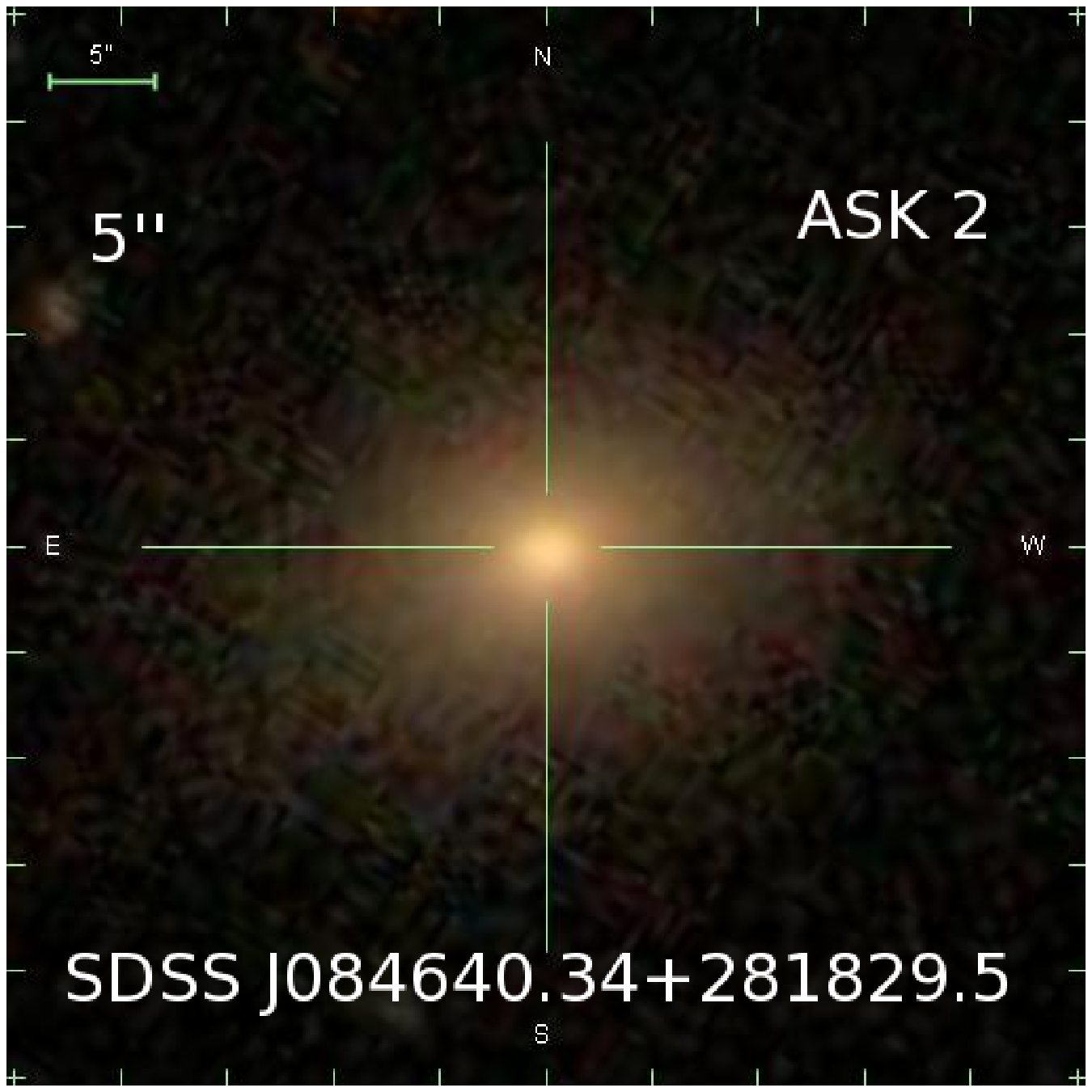}
\includegraphics[width=0.24\textwidth]{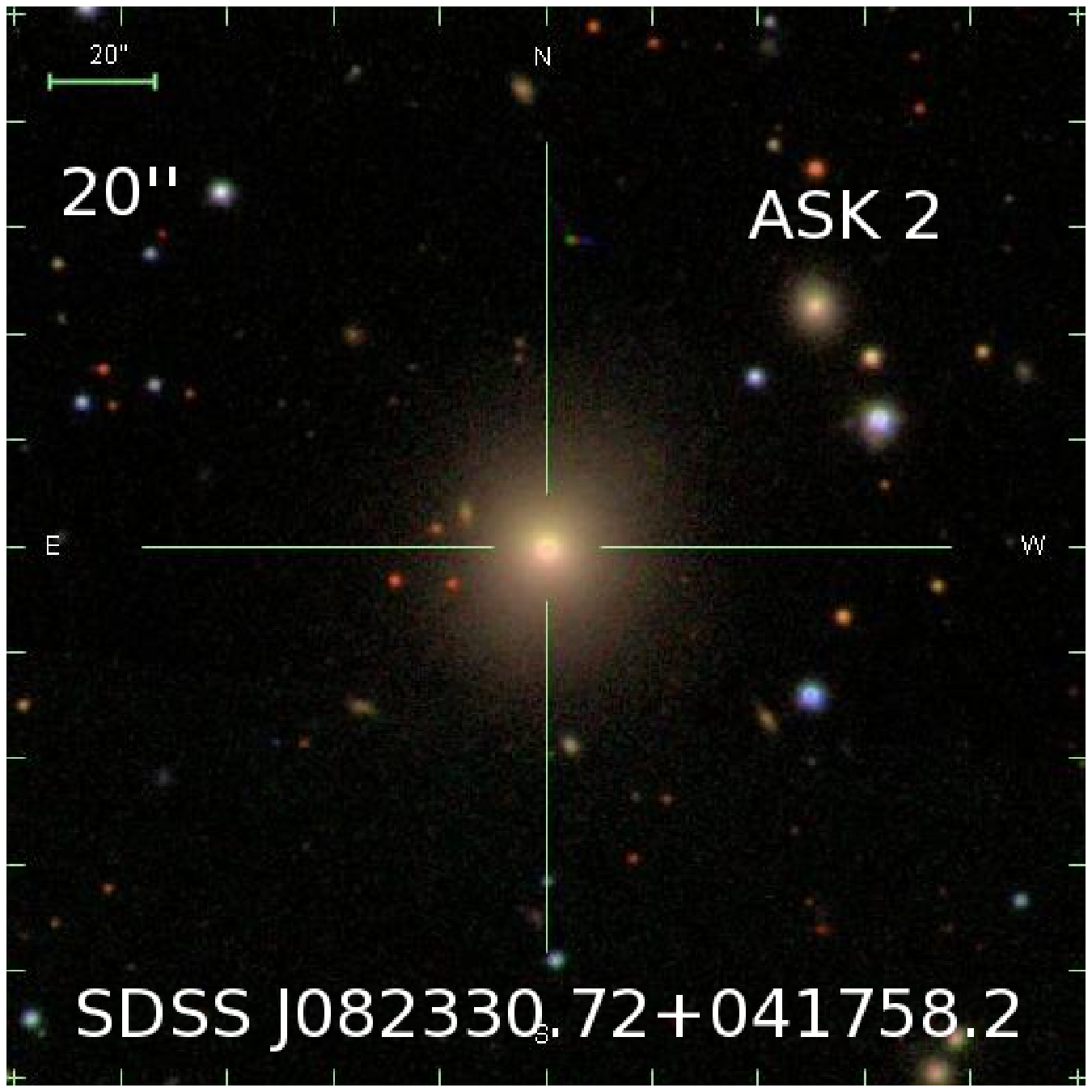}
\includegraphics[width=0.24\textwidth]{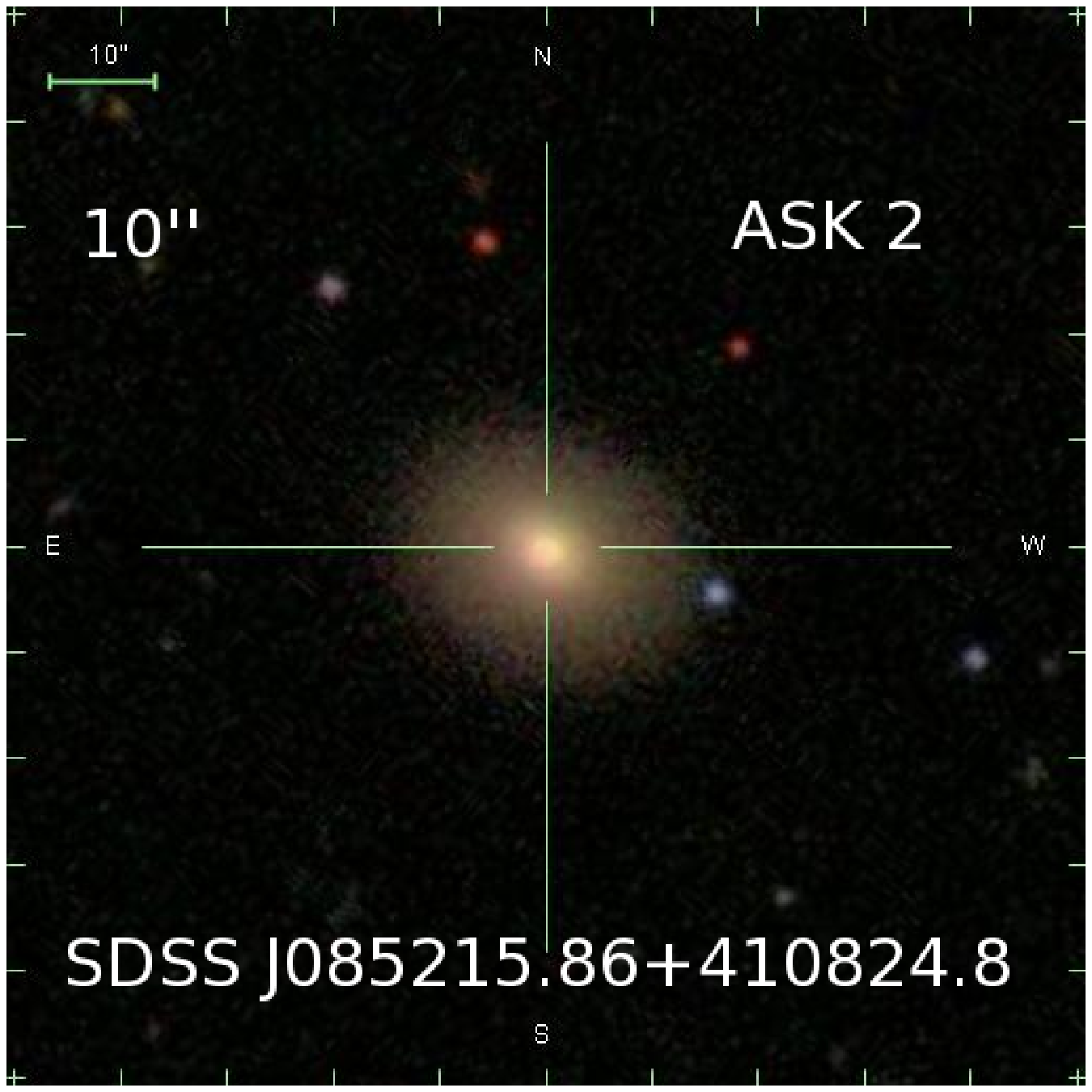}\\
\includegraphics[width=0.24\textwidth]{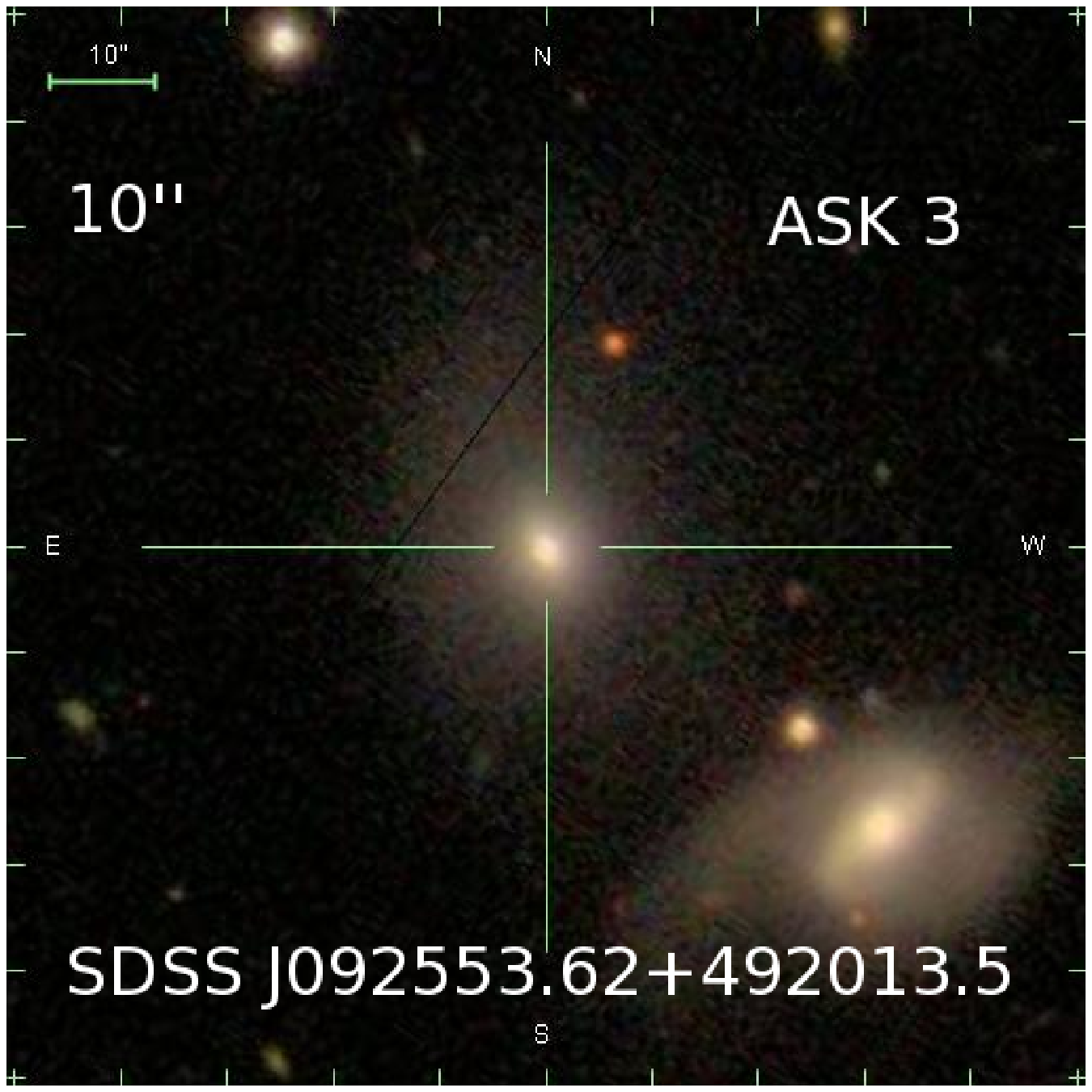}
\includegraphics[width=0.24\textwidth]{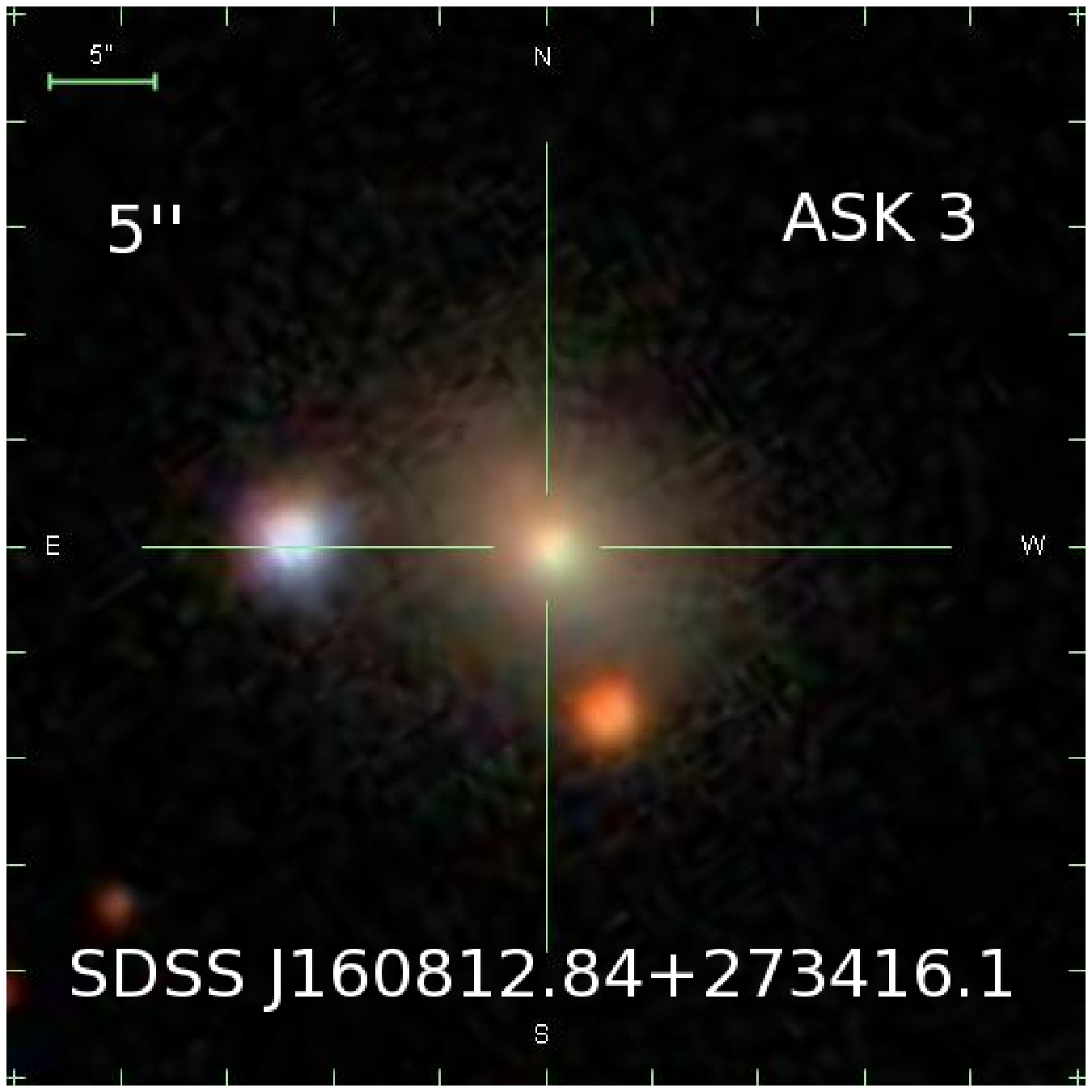}
\includegraphics[width=0.24\textwidth]{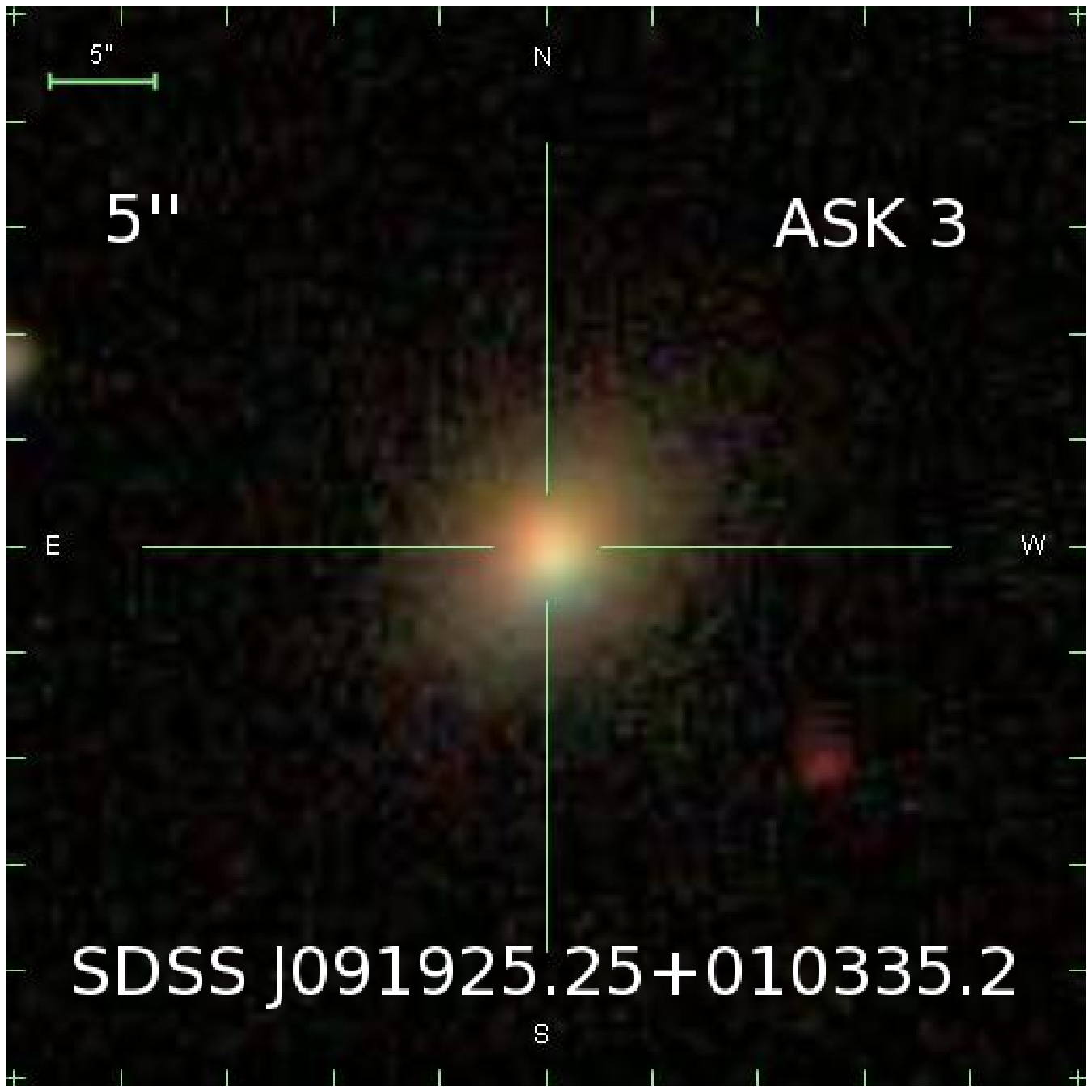}
\includegraphics[width=0.24\textwidth]{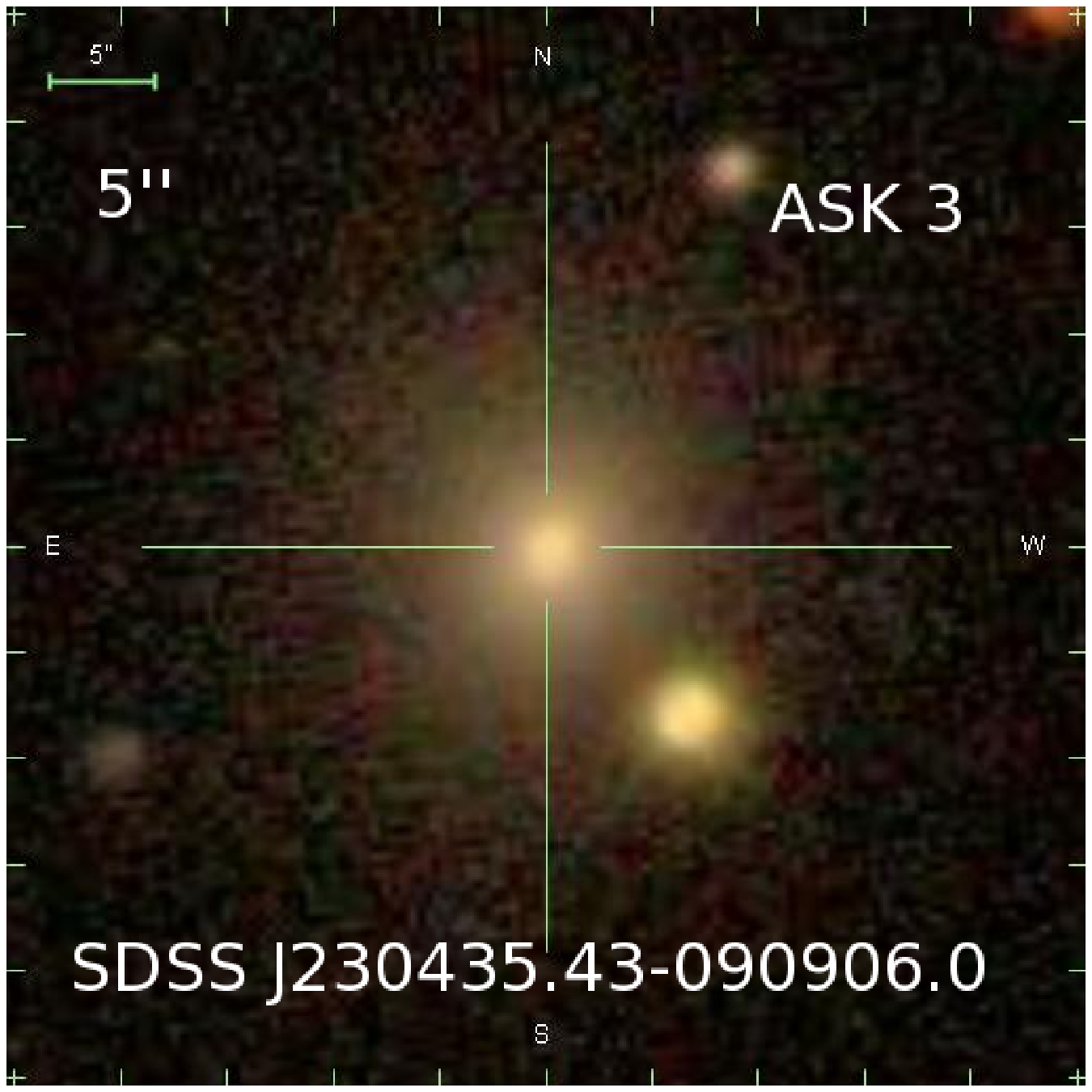}
\caption{Cutout images of  \rm{EASK0} (top row), \rm{EASK2} (middle row), and \rm{EASK3} (bottom row) galaxies with the highest weights.
}
\label{imagen}
\end{figure*}

   \begin{figure*}
   \centering
   \includegraphics[width=17cm,height=13cm]{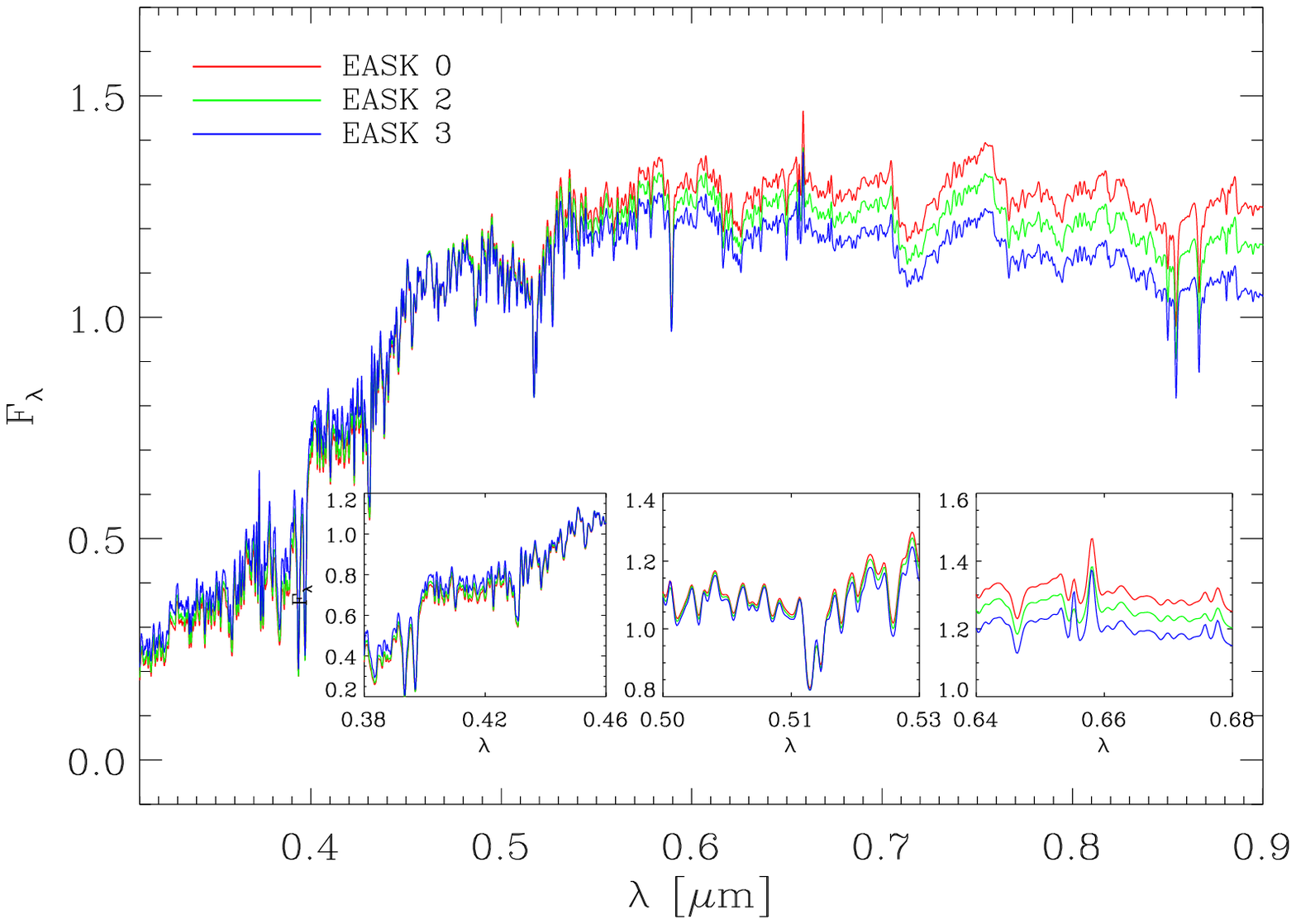}
   \caption{Template spectra of the EASK0 (red), EASK2 (green), and EASK3 (blue) spectroscopic classes. The small panels  zoom into of the spectral ranges of the 4000 $\rm{\AA}$ break (left), Mg absorption triplet (center), and H$\alpha$ (right). The fluxes of the three spectra are in dimensionless units. They were normalized to the average flux in the $g$-filter bandpass.}
              \label{fspec}%
    \end{figure*}

Galaxies classified morphologically as ellipticals do not belong to all ASK spectral classes. They are basically concentrated in the ASK classes called ASK0, ASK2, and ASK3 (see S\'anchez Almeida et al. 2011).  The concentration of elliptical galaxies in only three spectral classes is just because blue galaxies with elliptical morphology are very rare in the local universe (less than 10$\%$, see \cite{kannappan2009}, S\'anchez Almeida et al. 2011). Although, they become more important in the high redshift Universe (see \cite{huertas2010}). Taking into account this approximation, the weight assigned to each galaxy of the catalog is given by 

\begin{equation}
 P_{i}(\rm{E}) \approx \it{P}_{i,morph}(\rm{E})  \times (\it{P}_{i,\rm{ASK}}(A_{0})+\it{P}_{i,\rm{ASK}}(A_{2})+\it{P}_{i,\rm{ASK}}(A_{3})),
\end{equation}

\noindent
where $P_{i,\rm{ASK}}(A_{0})$, $P_{i,\rm{ASK}}(A_{1})$, and $P_{i,\rm{ASK}}(A_{3})$ are the probabilities of belonging to  ASK0, ASK2, and ASK3 spectroscopic classes, respectively.

In the present work, we have assigned three different weights to the family of elliptical galaxies corresponding to their three different main spectral types  (ASK0, ASK2, ASK3).  Hereafter we will call them \rm{EASK0}, \rm{EASK2}, and \rm{EASK3}, respectively. The values of the weights are given by,   

\begin{equation}
P_{i}(\rm{EASK0})=\it{P}_{i,morph}(\rm{E}) \times \it{P}_{i,\rm{ASK}}(A_{0}), 
\end{equation}
\begin{equation}
P_{i}(\rm{EASK2})=\it{P}_{i,morph}(\rm{E}) \times \it{P}_{i,\rm{ASK}}(A_{2}), 
\end{equation}
\begin{equation}
P_{i}(\rm{EASK3})=\it{P}_{i,morph}(\rm{E}) \times \it{P}_{i,\rm{ASK}}(A_{3}).
\end{equation}

 It is obvious from previous equations that: $P_{i}(\rm{E})=\it{P}_{i}(\rm{EASK0})+\it{P}_{i}(\rm{EASK2})+\it{P}_{i}(\rm{EASK3})$. In all relations analyzed  throughout the  paper, all galaxies contribute to each relation with the above weights.  Figure \ref{imagen} shows snapshots of the four galaxies with the highest $P_{i}(\rm{EASK0}), P_{i}(\rm{EASK2})$, and $P_{i}(\rm{EASK3})$ values.

In nature, we expect a gradual transition between the different spectroscopic and morphological galaxy types. The approach presented here is therefore adapted to this characteristics in the sense that each galaxy of the catalog belongs to the \rm{EASK0}, \rm{EASK2}, and \rm{EASK3} classes in proportion to its weight. This has the advantage that we have not introduced additional biases in the selection of the elliptical galaxy sample. Our set of galaxies has the same biases as the SDSS DR7 spectroscopic catalog.

Figure \ref{fspec} shows the mean spectra of the EASK0, EASK2, and EASK3 galaxies. These template spectra were obtained by combining the ASK0, ASK2 and ASK3 spectra weighted  as given in equations (2), (3) and (4), respectively. Notice  that the  three spectral classes correspond to galaxies showing typical spectra of early-type systems, i.e., well defined 4000\,\AA \  break, strong absorption Balmer lines, and no strong emission lines. Weak emission lines can be observed at H${\alpha}$, [NII], [SII] and [OII] wavelengths. Thus, galaxies from ASK0, ASK2 and ASK3 classes show H${\alpha}$ equivalent widths EW(H${\alpha}) <$ 1 \AA\ (see \cite{sanchezalmeida2010}). These weak emission lines in early-type galaxies have been observed before. Thus, up to 60$\%$ of the early-type galaxies show  EW(H${\alpha}) \ga$ 0.5 \AA\ (see, e.g., \cite{phillips1986}). 
Notice also that the emission of ASK0, ASK2 and ASK3 galaxies is in the region of the BPT diagram (\cite{baldwin1981}) 
traditionally attributed to AGNs and LINERs (\cite{sanchezalmeida2010}). 
However, this region of the diagram is also characteristics of 
ionization provided by old stellar populations  (\cite{binette1994, stasinska2008}), with the EW of H$\alpha$ being the parameter to distinguish 
the two possibilities  (\cite{cidfernandes2011}). 
Thus,  their weak emission lines, together with their position on the BPT diagram, seems to be consistent with EASK galaxies being retired galaxies where the interstellar medium is ionized by  hot low mass evolved stars. 

The ASK classification does not give a class of early-type galaxies without emission lines similar to 
those galaxies already observed (see e.g. \cite{bressan2006, annibali2010, panuzzo2011}). Since most of the early-type galaxies show weak emission lines (see e.g. \cite{phillips1986}), the presence of emission lines is  a feature bypassed by the automatic 
classification algorithm.
In order to separate them out, one would need to refine the classes
by re-running the  classification upon 
spectra of individual ASK classes (e.g., Morales-Luis et al. 2011),
but  such work has not been carried out yet.  

\subsection{Comparison with other samples of early-type galaxies}

In this section we quantify the contamination by spirals of our galaxy sample. We compare the apparent axis ratios distribution and the luminosity functions of our elliptical galaxies with other samples of visually classified ellipticals. In particular, we use as reference the recently published eyeball classification of 14034 galaxies by Nair \& Abraham (2010).

\subsubsection{Apparent axis ratio distribution}

 The intrinsic structure of galaxies is related to their apparent axial ratio ($b/a$). Elliptical and disc galaxies show very different axial ratio distribution. In particular, the axial ratio distribution of ellipticals peaks at $b/a\approx0.8$, and decreases to zero at $b/a\approx0.2$. In contrast, discs show an almost flat $b/a$ distribution in the range $b/a=0.2-0.8$ (e.g., \cite{tremblay1995}, \cite{padilla2008}).  In our approximation, the $b/a$ distribution of the elliptical galaxies is given by,

\begin{equation}
N(b/a) =\frac{1}{\Delta (b/a)}\sum_{i} P_{i}(\rm{E}) \ \Pi(\frac{(\it{b/a})_{i}-\it{b/a}}{\Delta (\it{b/a})}),
\end{equation}

\noindent
where the sum is over all the galaxies in the sample, $\Delta (b/a)$ represents the bin size of the histogram, $(b/a)_{i}$ is the aparent axial ratio of the i-th galaxy, and $\Pi$ is the rectangle function given by,

\begin{equation}
\Pi(x)= \left\{ \begin{array}{ll} 
1 & \mbox{if $|x|<1/2$}, \\
0 & \mbox{elsewhere}.
\end{array} \right. 
\end{equation}

Note that the histogram defined in Eq.(6) has an integral equal to $\sum_{i} P_{i}(\rm{E})$.

\begin{figure}
   \centering
   \includegraphics[width=9cm,height=7cm]{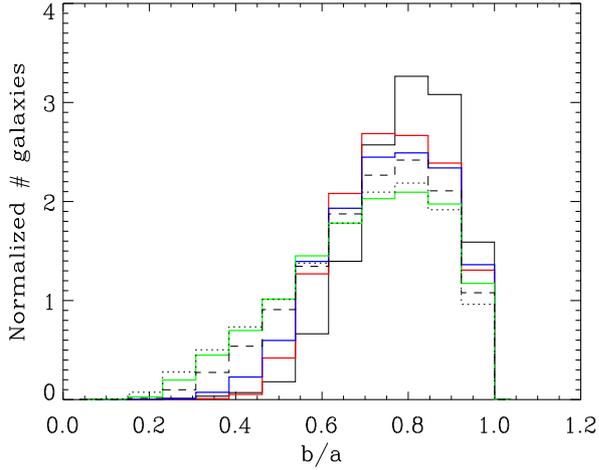}
   \caption{Axial ratio distribution of elliptical galaxies as computed in this work (black full line). The axial ratio distributions of galaxies visually classified by Nair \& Abraham (2010) as Elliptical (red line), Elliptical + S0 (blue line), and Elliptical + S0+Sab (green line) are also overplotted. The dashed and dotted lines represent the axial ratios of early-type galaxies from Bernardi et al. (2003) and Strateva et al. (2001), respectively. All histograms have been normalized to the same area.}
              \label{felip}%
    \end{figure}

Figure \ref{felip} shows this axial ratio distribution. We have also overplotted  the axial ratio distributions of those galaxies morphologically classified as E, E+S0, and E+S0+Sab  by Nair \& Abraham (2010). Notice that the $b/a$ distributions of E, E+S0 and E+S0+Sab all peak at about $b/a\sim0.8$. Nevertheless, the inclusion of discs changes the tail of the distribution.  Thus, discs increase the fraction of galaxies with $b/a<0.5$. In  Nair \& Abraham sample, 2\%, 4\%, and 43\% of the E, E+S0, and E+S0+Sab galaxies show $b/a<0.5$. In our selected E sample, only 1\% of the galaxies have $b/a<0.5$. This indicates that the tail of the $b/a$ distribution of our E sample is closer to the tail of the axial ratio distribution of the visually classified ellipticals. Then, we can conclude that our automated galaxy classification  is not significantly contaminated by discs, which would appear as an extended tail towards small $b/a$. 

We have also overplotted in Fig.\ref{felip} the axial ratio distribution of the early-type galaxies selected in Strateva et al. (2001) and Bernardi et al. (2003). As it is revealed by  the tail of the distributions, both samples are contaminated by disc galaxies, being the color selected sample (\cite{strateva2001}) the most contaminated (see \cite{bernardi2010} for a discussion about biases in selection of early-type galaxies).

\begin{figure}
   \centering
   \includegraphics[width=9cm,height=7cm]{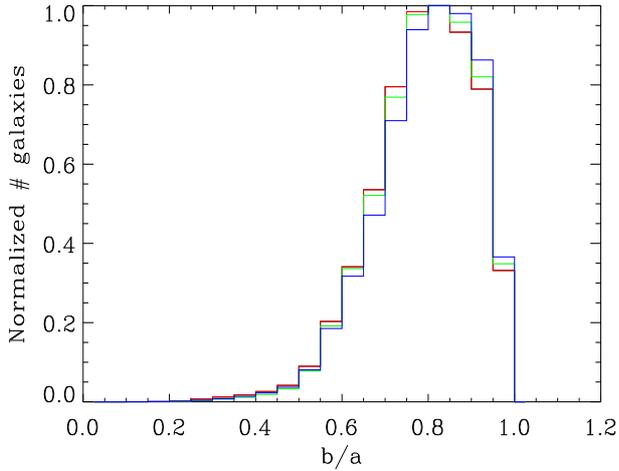}
   \caption{Axial ratio distributions of \rm{EASK0} (red), \rm{EASK2}(green), and \rm{EASK3} (blue) galaxies. All the histograms are normalized to their peaks.}
              \label{felipclas}%
    \end{figure}

The $b/a$ distributions of \rm{EASK0}, \rm{EASK2}, and \rm{EASK3} galaxies can be computed by substituting in eq. (5) $P_{i}(\rm{E})$ with $P_{i}(\rm{EASK0}) , \it{P}_{i}(\rm{EASK2})$, and $P_{i}(\rm{EASK3})$, respectively. These $b/a$ distributions are shown in Fig. \ref{felipclas}. We do not find systematic differences between the different spectral classes of galaxies and the global elliptical galaxy sample. This fact also indicates that if there is some disc contamination left, it is the same for the three classes of elliptical galaxies defined in the present work.


\subsubsection{Luminosity function}

The galaxy luminosity function, $\Phi(L)$, is one of the fundamental statistics in galaxy studies. It measures the number of galaxies in  a given volume per bin of luminosity. In magnitude limited samples, it computations requires considering the fact that brighter galaxies can be seen further away. This is done by computing $V_{max}$, the maximum volume within which a galaxy of a given magnitude could have been detected. Each galaxy is then weighted by $1/V_{max}$. Thus, the number of elliptical galaxies per unit magnitude and volume is given by,

\begin{equation}
\Phi(M)=\frac{1}{\Delta M}\sum_{i} \frac{P_{i}(\rm{E})}{V_{i, max}} \ \Pi(\frac{M_{i}-M}{\Delta M}),
\end{equation}

\noindent
where the sum comprises all the galaxies of the sample, $\Delta M$ represents magnitude bin size, and $V_{i,max}$ is the maximum volume of the $\rm{i-th}$ galaxy.  In a similar way, we can compute the luminosity functions of \rm{EASK0}, \rm{EASK2}, and \rm{EASK3} galaxies by replacing $P_{i}(\rm{E})$ with $P_{i}(\rm{ASK0})$, $P_{i}(\rm{ASK2})$, or $P_{i}(\rm{ASK3})$.

\begin{figure*}
   \centering
   \includegraphics[width=17cm,height=13cm]{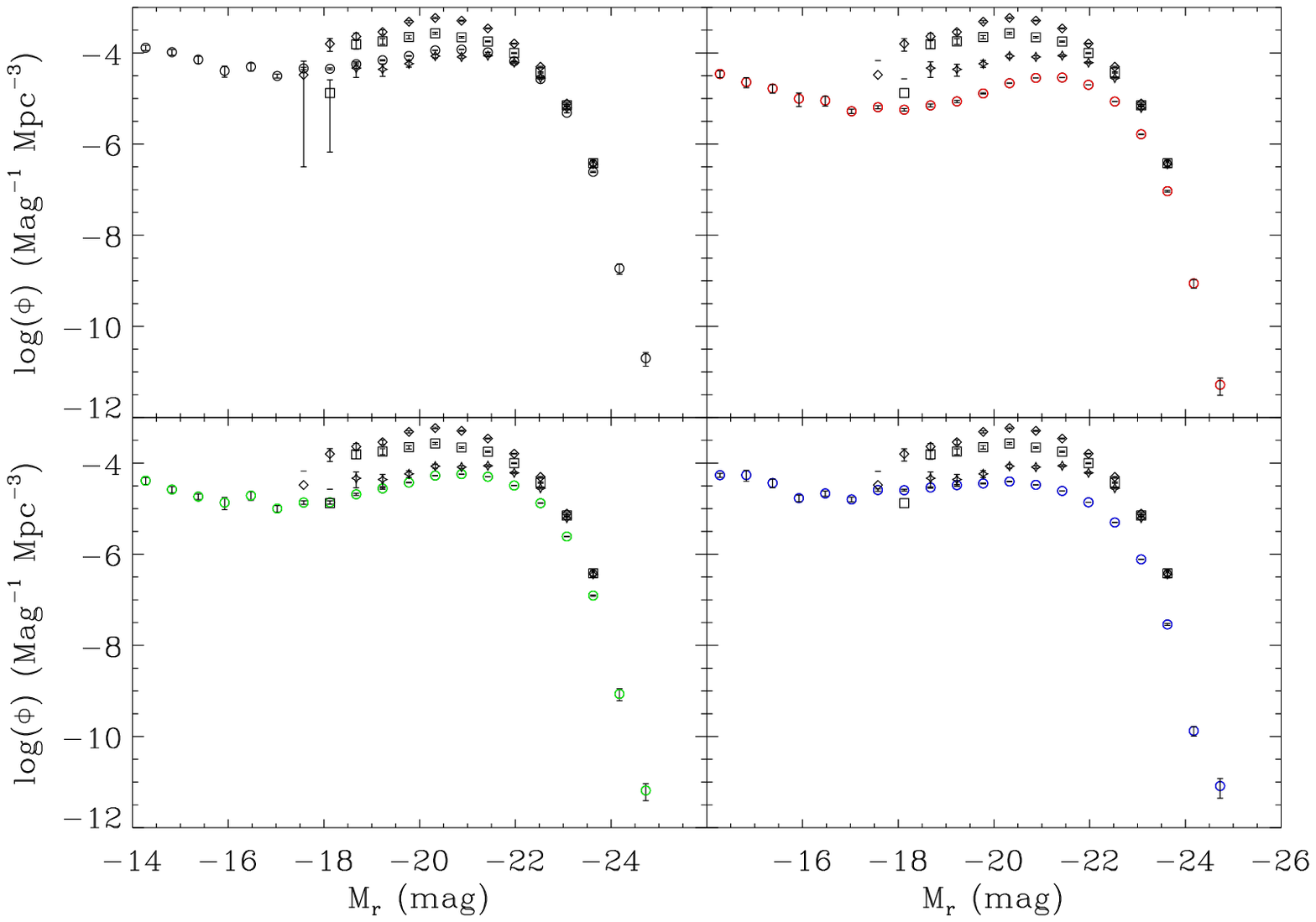}
   \caption{Luminosity functions of our elliptical galaxies (top left, black open circles), \rm{EASK0} (top right, red open circles), \rm{EASK2} (bottom left, green open circles), and \rm{EASK3} (bottom right, blue open circles). In all panels we represent as reference the luminosity functions of E (open stars), E+S0 (open squares) and E+S0+Sa (open diamonds) visually classified by Nair \& Abraham (2010). The errors of the luminosity functions were computed by bootstraping.}
              \label{lfcomp}%
    \end{figure*}

Figure \ref{lfcomp} shows the luminosity functions (LF) of our elliptical selected galaxies, and those from \rm{EASK0}, \rm{EASK2}, and \rm{EASK3} galaxies. 
We have also overplotted  the LF of those galaxies visually selected as E, E+S0, and E+S0+Sa by Nair \& Abraham (2010). 

Notice that the LF of our EASK0, EASK2, EASK3, all combined, follows the LF of the visually identified  ellipticals (top left panel of Fig.~\ref{lfcomp}). In addition, Fig. \ref{lfcomp} shows that EASK2 galaxies dominate the LF of our ellipticals at all absolute magnitudes. The LF of EASK3 galaxies are close to the LF of visually selected E galaxies at the low magnitude regime $M_{r}<-18$ (see also Sec. 4).

These facts also support that our rendering of elliptical galaxies is not contaminated by discs.

   \begin{figure}
   \centering
   \includegraphics[width=9cm,height=7cm]{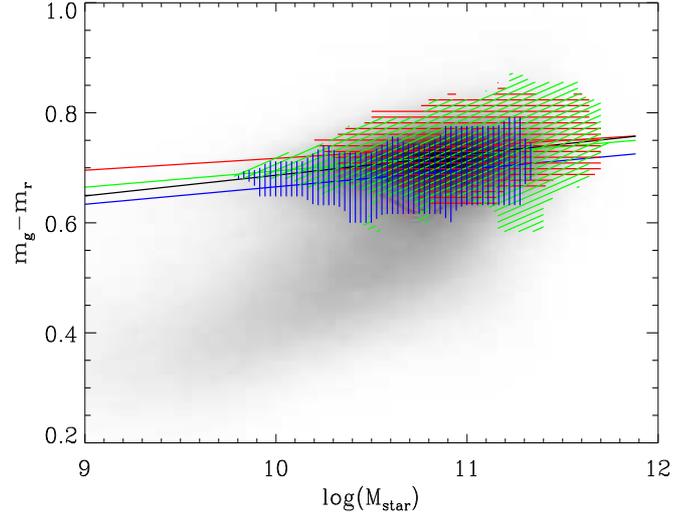}
   \caption{Color-mass diagram of the galaxies from SDSS-DR7 spectroscopic sample (grey scale). Colored regions show the location of galaxies with $P_{i}(\rm{EASK0})>0.5$ (red), $P_{i}(\rm{EASK2})>0.5$ (green), and $P_{i}(\rm{EASK3})>0.5$ (blue). The red, green, blue and black solid lines represent linear fits to the color-mass relations weighted with $P_{i}(\rm{EASK0}), P_{i}(\rm{EASK2}), P_{i}(\rm{EASK3})$, and $P_{i}(\rm{E})$, respectively.}
              \label{f1}%
    \end{figure}

\section{Scaling relations of elliptical galaxies}

In this section we present the scaling properties of the galaxies  \rm{EASK0}, \rm{EASK2}, and \rm{EASK3}. The relations were derived including all galaxies but weighted according to $P_{i}(\rm{EASK0})$, $P_{i}(\rm{EASK2})$ and $P_{i}(\rm{EASK3})$, respectively (see Sect. 2). 

\subsection{Color-mass diagram}

Figure \ref{f1} shows the color-mass diagram of all the galaxies in the  SDSS-DR7 spectroscopic sample. The $m_{g}-m_{r}$ color of the galaxies was obtained from the petrosian $g$ and $r$ magnitudes downloaded from the SDSS DR7 database. These magnitudes were amendded to account for Galactic dust attenuation and k-correction. Galactic absorptions and k-corrections were computed from the MPA-JHU DR7 spectrum measurements\footnote{http://www.mpa-garching.mpg.de/SDSS/DR7} by subtracting the fiber magnitude and the derredening and de-deredshifted magnitudes of each galaxy (\emph{plug\_mag} and \emph{kor\_mag}). Stellar masses were also obtained from the MPA-JHU DR7 release. They were computed following the strategy in Kaufmann et al. (2003). We have also overplotted  in Fig. \ref{f1} the regions occupied by galaxies with $P_{i}(\rm{EASK0})>0.5$, $P_{i}(\rm{EASK2})>0.5$, and $P_{i}(\rm{EASK3})>0.5$. These areas enclose the galaxies representative of the three spectral classes. Note that these galaxies populate the red-sequence of the color-mass diagram. 

It can also be seen in Fig. \ref{f1} that the mass range of the three classes of galaxies is different. Thus, galaxies with $P_{i}(\rm{EASK0})>0.5$ span the range $10.2<\log(M_{star})<11.8$, whereas those with $P_{i}(\rm{EASK3})>0.5$ have $9.8<\log(M_{star})<11.3$. In contrast, galaxies with $P_{i}(\rm{EASK2})>0.5$ occupy the full mass range ($9.8<\log(M_{star}<12$). We refer to  Section \ref{Sec:discussion} for a detailed dicussion on the mass distribution of the different classes.

We have performed linear fits  $m_{g}-m_{r}=a+b \log(M_{star})$ to the color-mass relation using $P_{i}(\rm{E})$, $P_{i}(\rm{EASK0})$, $P_{i}(\rm{EASK2})$ and $P_{i}(\rm{EASK3})$ as weights. The best fit coefficients are given in Tab. \ref{tab3}. Notice that for a given stellar mass, galaxies from \rm{EASK3} class are bluer on average.

\subsection{Mass-size relation}
   \begin{figure}
   \centering
   \includegraphics[width=9cm,height=7cm]{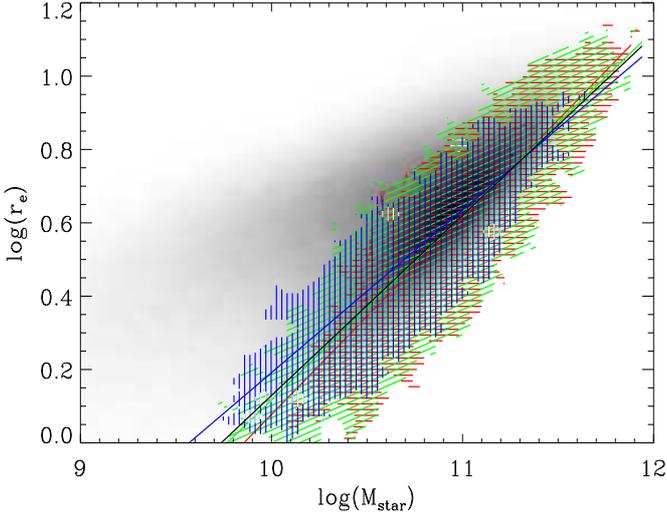}
   \caption{Effective r-band radius ($r_{e})$ as a function of stellar mass for galaxies in the SDSS DR7 spectroscopic catalog (grey scale). The colored regions show the location of galaxies with $P_{i}(\rm{EASK0})>0.5$ (red), $P_{i}(\rm{EASK2})>0.5$ (green), and $P_{i}(\rm{EASK3})>0.5$ (blue). The solid lines show the best linear weighted fits to the relations (see Fig. \ref{f1} for the color code of these lines).}
              \label{f2}%
    \end{figure}

According to the virial theorem, masses and sizes of galaxies are well correlated. Basically, more massive (or luminous) galaxies are larger than their low mass counterparts. This correlation is the so-called mass/luminosity-size relation, and it is followed by galaxies in all range of luminosities and morphological types. However, there are important differences depending on the  morphological type. In particular, the mass-size relation of early-type galaxies is less scattered than the one of  late-type systems, even if there is a strong dependence with luminosity (see \cite{shen2003}). Figure \ref{f2} shows the relation between the effective radius ($r_{e}$) and the stellar mass of the galaxies in our sample. We have taken the radius containing  $50\%$ of the total petrosian galaxy luminosity as an estimate for $r_{e}$. Figure \ref{f2} also shows in colors the location of the galaxies with $P_{i}(\rm{EASK0})$, $P_{i}(\rm{EASK2})$, and $P_{i}(\rm{EASK3})$ larger than 0.5.  Note that the size-mass relation steepens for ellipticals as compared to the full set of galaxies, and so, as compared to disc galaxies.  We have also performed linear fits to the size-mass relation using $P_{i}(\rm{EASK0})$, $P_{i}(\rm{EASK2})$, and $P_{i}(\rm{EASK3})$ as weights. They are shown in Fig. \ref{f2}. The coefficients of the best fit $\log(r_{e})=a+b \log(M_{star})$  are listed in Tab.\ref{tab3}. Notice that we also see deviations from a pure straight line at the low and high mass ends reported in previous works (see \cite{bernardi2010}).

\begin{table}
\caption{Coefficients of linear fits to the color-mass, size-mass, Faber-Jackson and FP relations for \rm{EASK0}, \rm{EASK2}, \rm{EASK3}, and as well as for all of them combined.}             
\centering        
\begin{tabular}{c c c c} 
\hline\hline                 
Relation & Galaxy class & $a$ & $b$ \\  
\hline     
 Color-Mass &  \rm{EASK0} & 0.51$\pm$0.46 & 0.02$\pm$0.04  \\      
   & \rm{EASK2} & 0.41$\pm$0.26 & 0.03$\pm$0.02      \\
   & \rm{EASK3} & 0.36$\pm$0.41 & 0.03$\pm$0.04     \\
  & All     &  0.32$\pm$0.13 & 0.04$\pm$0.01  \\
\hline           
   Mass-size &\rm{EASK0} & -5.29$\pm$0.42 & 0.54$\pm$0.04  \\      
   & \rm{EASK2} & -4.92$\pm$0.26 & 0.50$\pm$0.02      \\
   & \rm{EASK3} & -4.25$\pm$0.43 & 0.44$\pm$0.04     \\
   & All & -4.78$\pm$0.13 & 0.49$\pm$0.01 \\
\hline
Faber-Jackson & \rm{EASK0} & -0.49$\pm$0.45& 0.25$\pm$0.04\\
& \rm{EASK2} & -0.72$\pm$0.28& 0.27$\pm$0.03\\
& \rm{EASK3} & -0.95$\pm$0.45&0.29$\pm$0.04 \\
& All & -0.81$\pm$0.14 & 0.28$\pm$0.01 \\
\hline
Fundamental Plane & \rm{EASK0} & 3.65$\pm$0.26 & 0.65$\pm$0.05 \\
& \rm{EASK2} & 3.65$\pm$0.17 & 0.64$\pm$0.04 \\
& \rm{EASK3} & 3.37$\pm$0.31 & 0.58$\pm$0.06 \\
& All & 3.57$\pm$0.09 & 0.63$\pm$0.02 \\ 
\hline                                   
\end{tabular}
\label{tab3}
\end{table}

Galaxies in the \rm{EASK0} and \rm{EASK2} classes show similar size-mass relations. In contrast, galaxies from the \rm{EASK3} class follow a size-mass relation with a slightly different slope. In particular, for a given stellar mass, galaxies from the \rm{EASK3} class are larger. The relation obtained for all the 3 classes together (using $P(\rm{E})$ as weight) is consistent with the one reported by Hyde \& Bernardi (2009) for their early-type galaxy sample ($\log(r_{e})=(-4.79\pm0.02) + (0.489\pm0.002) \log(M_{star})$).

\subsection{The Fundamental Plane}

   \begin{figure}
   \centering
   \includegraphics[width=9cm,height=7cm]{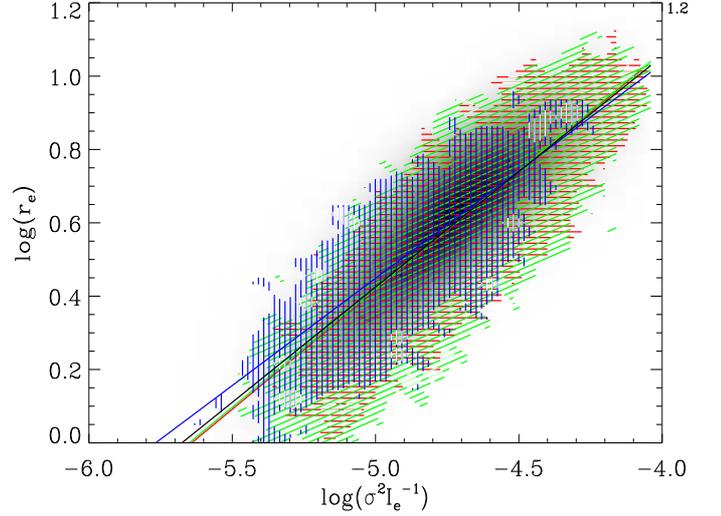}
   \caption{Fundamental plane of all galaxies from the SDSS-DR7 spectroscopic catalog (grey scale). The colored regions show the location of galaxies with $P_{i}(\rm{EASK0})>0.5$ (red), $P_{i}(\rm{EASK2})>0.5$ (green), and $P_{i}(\rm{EASK3})>0.5$ (blue). The continuous lines show the best linear weighted fits to the relations (colors of the solid lines as in Fig.\ref{f1}).}
              \label{f4}%
    \end{figure}

Galaxies are not located randomly in the space defined by the effective radius ($r_{e}$), velocity dispersion ($\sigma$), and mean effective surface brightness ($I_{e}$). They follow a tight relationship called fundamental plane (FP; \cite{djorgovsky1987}, \cite{dressler1987}) given by,

\begin{equation}
r_{e} \propto \sigma^{\alpha} I_{e}^{-\beta}.
\end{equation}

 This relation is a consequence of the dynamical equilibrium (virial theorem) together with the regular behavior of both mass-luminosity ratio and structure of early-type galaxies. According to the virial theorem, $\alpha=2$ and $\beta=1$. There are, however, differences between these theoretical values and those obtained from observations. These discrepancies are called the tilt of the FP, and imply that the mass-to-light ratio is a function of galaxy mass or luminosity ($M/L \propto L^{\gamma}$).

We have adopted the approach of plotting the FP in the $r_{e}$ vs $\sigma^{2}I_{e}^{-1}$ virial plane coordinates  (see e.g. \cite{robertson2006}). In this representation the tilt of the FP can be quantified by the relation

\begin{equation}
r_{e}\propto (\sigma^{2}I_{e}^{-1})^{\lambda},
\end{equation}

\noindent
where $\lambda=1$ means an alignment of the FP with the virial expectation.

We use the velocity dispersion at the effective radius ($\sigma_{r_{e}}$) as an estimate of the central velocity dispersion ($\sigma$). Aperture corrections were however applied to the values given in the SDSS catalog. These corrections depend on the galaxy type since the shape of the velocity dispersion radial profile depends on morphology. As a matter of fact, early-type galaxies present important radial gradients in their velocity dispersion while late-type systems show almost flat profiles. Following our weight-based solution, the aperture correction applied to the velocity dispersion of each galaxy ($\sigma_{i,r_{e}})$ is given by,

\begin{equation}
\begin{array}{l}
\sigma_{i,r_{e}}^2=(\sigma^{2}_{i,cor}\times(P_{i}(\rm{E})+\it{P}_{i}(\rm{S0})+\it{P}_{i}(\rm{Sab})) + \\
           \ \ \ \ \ \ \ \ \ \ \ \ \ \sigma^{2}_{i,SDSS}\times\it{P}_{i}(\rm{Scd}))),
\end{array}
\label{eq:sigmas}
\end{equation}

\noindent
where $\sigma_{i,SDSS}$ is the velocity dispersion of the i-th galaxy given in the SDSS-DR7 catalog\footnote{The velocity dispersion of the galaxies was corrected for the spectral resolution of SDSS}, $P_{i}(\rm{E}), \it{P}_{i}(\rm{S0}), \it{P}_{i}(\rm{Sab})$ and $P_{i}(\rm{Scd})$ are the probabilities of each galaxy to be classified as elliptical, S0, Sab and Scd, respectively (see \cite{huertas2011}), and $\sigma_{i,cor}$ is the aperture correction for early-type galaxies given by Jorgensen et al. (1995):  $\sigma_{i,cor}=\sigma_{i,SDSS}\times(\frac{r_{ap}}{r_{e}})^{0.04}$, with $r_{ap}=1.5$ arcsec corresponding to the radius of the SDSS fiber. When $P(\rm{E}), \it{P}(\rm{S0})$ or $P(\rm{Sab})$  are large, then the computed $\sigma$ is close to the one worked out by Jorgensen et al. (1995). No aperture correction is taken into account for galaxies with a large probability of being Scd. 

Figure \ref{f4} shows the distribution of all galaxies  in the $\log(r_{e})$ vs $\log(\sigma^{2}I_{e}^{-1})$ plane. We have also overplotted the location of galaxies with $P_{i}(\rm{EASK0}), \it{P}_{i}(\rm{EASK2})$, and $P_{i}(\rm{EASK3})$ greater than 0.5. Figure \ref{f4} also shows the linear fits $\log(r_{e})=a+b\log(\sigma^{2}I_{e}^{-1})$ using these probabilities as weights. The coefficients of these relations are given in Tab.\ref{tab3}. Notice that all classes present FPs with tilt ($b\neq 1$). The tilt is however larger for \rm{EASK3} galaxies.

\subsection{The Faber-Jackson relation}

\begin{figure}
   \centering
   \includegraphics[width=9cm,height=7cm]{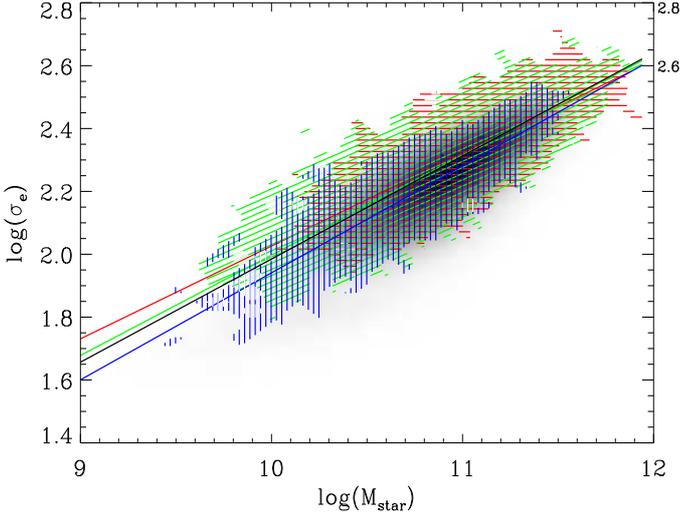}
   \caption{The Faber-Jackson relationship for the galaxies in the SDSS-DR7 spectroscopic catalog (grey scale). The colored regions show the location of galaxies with $P_{i}(\rm{EASK0})>0.5$ (red), $P_{i}(\rm{EASK2})>0.5$ (green), and $P_{i}(\rm{EASK3})>0.5$ (blue). The solid lines correspond to the best linear weighted fits to the relations (colors of the solid lines as in Fig. \ref{f1}).}
              \label{f3}%
    \end{figure}

The Faber-Jackson relation (FJR; \cite{faber1976}) is one of the most studied projections of the fundamental plane. It shows a tight correlation between mass (luminosity) and central velocity dispersion for early-type galaxies, i.e. more massive (luminous) galaxies have larger central velocity dispersions. Figure \ref{f3} shows the FJR for all galaxies in our sample. 

 Like in previous figures, we also show the locations of galaxies with $P_{i}(\rm{E})$, $P_{i}(\rm{EASK0})$, $P_{i}(\rm{EASK2})$, and $P_{i}(\rm{EASK3})$ greater than 0.5 and the best weighted linear fits $\log(\sigma)=a+b \log(M_{star})$. Coefficients of the best fits  are given in Tab. \ref{tab3}. Again, galaxies from \rm{EASK0} and \rm{EASK2} classes show similar FJRs. However, galaxies from \rm{EASK3} show a different relation. Thus, for a fixed stellar mass, galaxies from \rm{EASK3} show smaller velocity dispersion than those from \rm{EASK0} and \rm{EASK2} classes. We have also fitted the FJR for all the elliptical sample (see Tab.\ref{tab3}). This relation is also consistent with the one reported by Hyde \& Bernardi (2009) for their early-type galaxy sample ($\log(\sigma)=(-0.86\pm0.02) + (0.286\pm0.002) \log(M_{star})$).

\section{Discussion}

\label{Sec:discussion}

\begin{figure}
   \centering
   \includegraphics[width=9cm,height=7cm]{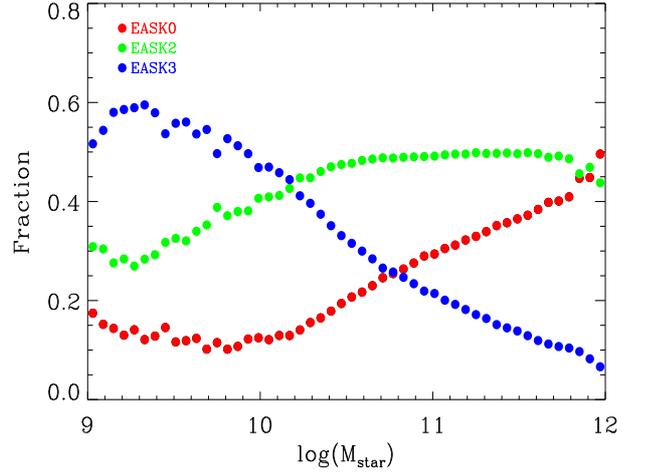}
   \caption{Mass fraction of \rm{EASK0} (red points), \rm{EASK2} (green points) and \rm{EASK3} (blue points) galaxies in each stellar mass bin.}
              \label{mf}%
    \end{figure}

In the previous section we have derived the scaling relations of the 3 spectral classes defining our elliptical sample. Before interpreting them, we compute the total weight of \rm{EASK0}, \rm{EASK2}, and \rm{EASK3} galaxies. We find that  28$\%$ of the total weight correspond to \rm{EASK0}, 49$\%$ to \rm{EASK2}, and 23$\%$ to \rm{EASK3}. \rm{EASK2} galaxies therefore represent half of the total sample. We also show in Fig. \ref{mf}  the mass fraction (MF) of the 3 spectro-morphological classes. The MF of a given class (EASKi; i=0,2, or 3)  and stellar mass ($M_{star}$) was obtained with the following expression: 

\begin{equation}
\rm{MF}(\rm{EASKi}; M_{star})=\frac{\sum_{j}\it{P}_{j}(\rm{EASKi}) \ \Pi(\frac{\it{M}_{star,j}-\it{M}_{star}}{\Delta \it{M}_{star}})}{\sum_{j}(\it{P}_{j}(\rm{E})}
\end{equation}

\noindent
where the sum is over all the galaxies of the sample of a given $M_{star}$, and $\Delta M$ is the stellar mass bin. Clearly, the \rm{EASK2} class dominates in a wide range of stellar masses (10.2$< \log(M_{star}) <11.7$). \rm{EASK3} becomes dominant in the low mass-end ($\log(M_{star})<10$) while \rm{EASK0} galaxies tend to be more abundant at high stellar masses. In the following we will therefore consider the \rm{EASK2} class as the canonical class of elliptical galaxies, and interpret the relative differences seen in the two other classes.

We have also studied the star formation histories (SFH) and metallicities of each spectro-morphological class.
 This was done as an independent confirmation of the results based on the spectral indexes. 
This was done using the data from Cid-Fernandes et al. 
(2005). They obtained the SFH of galaxies in SDSS-DR7 spectroscopic catalog by using STARLIGHT, a code which fits the observed spectrum of each galaxy by a combination of single stellar populations (SSP) spectra of different ages and metallicities. Thus, the model spectrum is given by:
\begin{equation}
M_{\lambda}=M_{\lambda_{0}}~(\Sigma_{j=1}^{N_{*}} x_{j}b_{j,\lambda}r_{\lambda}) \,\otimes\, G(v_{*},\sigma_{*})
\end{equation}
where $b_{j,\lambda}$ is the spectrum of the $j$-th SSP normalized at $\lambda_{0}$, $M_{\lambda_{0}}$ is the 
synthetic flux at the  normalization wavelength, $N_{*}$ is the number of SSPs used in the fit,  $x_j$ is the population vector, and the symbol $\otimes$ denotes convolution. The extinction, $r_\lambda$, assumes  a galaxy-like law  with a single free parameter (\cite{cardelli1989}).
The modelled spectrum $M_{\lambda}$ is convolved with a Gaussian distribution, $G(v_{*},\sigma_{*})$, centered at velocity $v_{*}$ and with  dispersion $\sigma_{*}$. Since all SSPs are assumed to start with the same mass, the components of the population vector $x_{j}$ represent the fractional mass contribution  of each SSP to the model flux at $\lambda_{0}$. For the SDSS-DR7 galaxies, Cid-Fernandes et al. used 150 single stellar population templates based on Bruzual \& Charlot (2003) models combined according to {\em Padova 1994} evolutionary tracks 
(Girardi 1996, and references therein). They cover a grid of 6 metallicities
 (from 0.005 to 2.5 times solar) and 25 ages 
(from 1 Myr to 18 Gyr). Further details are given in Sect.~2.1 of Asari et al.~(2007).  The Code searches for the 
minimum $\chi^{2}=\Sigma_{\lambda} ((O_{\lambda}-M_{\lambda})\,w_{\lambda})^{2}$, where $O_{\lambda}$ is the 
observed spectrum and $w_{\lambda}^{-1}$ its errors. STARLIGHT uses the Metropolis scheme for the 
$\chi^{2}$ minimization (see \cite{cidfernandes2005} for a full description of the code).

The population vector $x_j$ of each galaxy in the SDSS sample was computed by Cid Fernandes 
(2010, private communication).
We use them to characterize   mean SFHs and metallicities.

\begin{figure*}
   \centering
   \includegraphics[width=17cm,height=12cm]{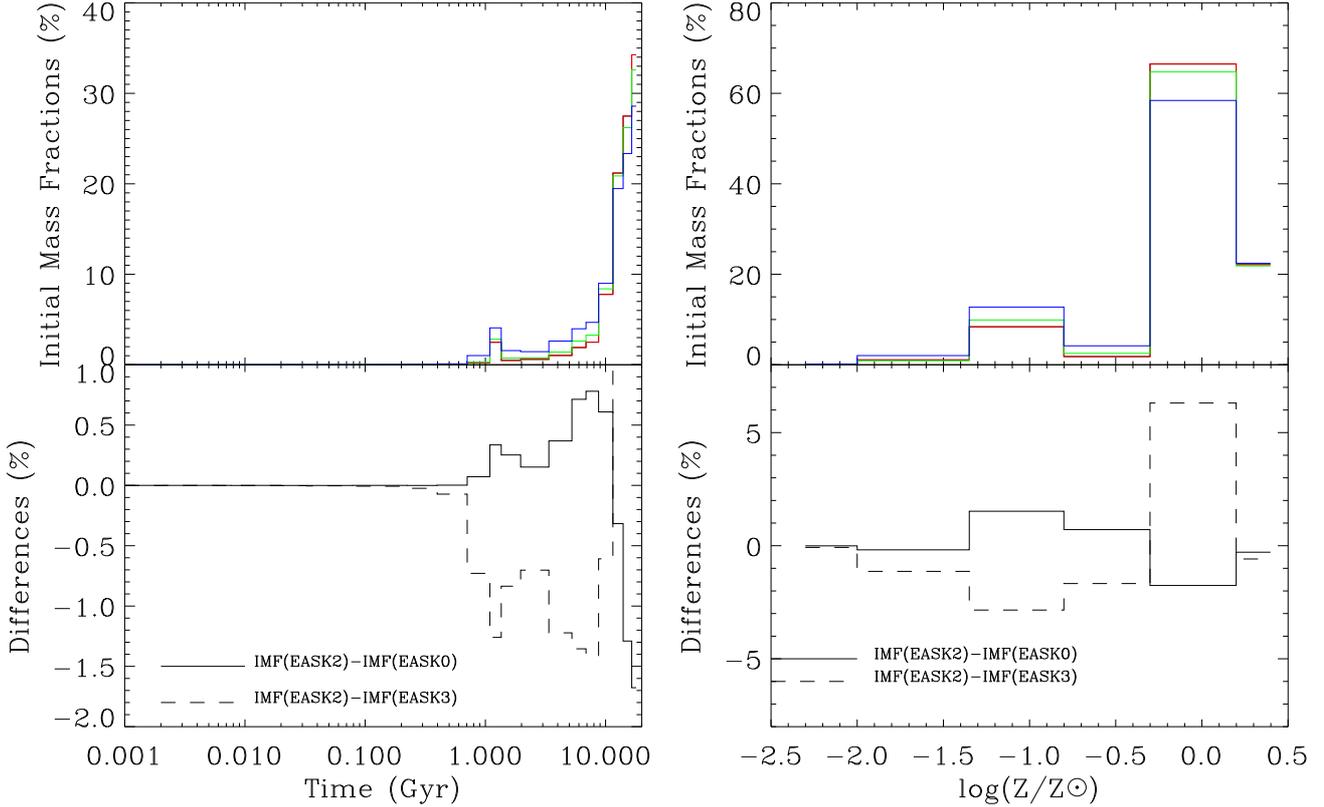}
   \caption{Top row: Mean star formation history (left pannel) and metallicities (right pannel) for \rm{EASK0} (red), \rm{EASK2} (green) and \rm{EASK3} (blue). Bottom row: Differences of the SFH (left pannel) and metallicities (right pannel) with respect to \rm{EASK2} galaxies.}
              \label{fig:SFH}%
    \end{figure*}

Like in previous sections, the mean SFH and metallicities of each spectro-morphological class of ellipticals were obtained by using $P_{i}(\rm{EASK0})$, $P_{i}(\rm{EASK2})$, and $P_{i}(\rm{EASK3})$ as weights. Figure\ref{fig:SFH} shows the results as the percentage of stellar mass created at each time, $x_j$, and with each metallicity, Z. 

Details are discussed below, but note that most of the stars were created long ago (time $>$ 10 Gyr) and 
with high metallicity (solar or larger).

\subsection{\rm{EASK2}: canonical class of elliptical galaxies}

\rm{EASK2} galaxies represent  50$\%$ of all elliptical galaxies in our sample. Moreover, they are well defined as spectral class according to the ASK classification (see Fig. 8 from \cite{sanchezalmeida2010}) in the sense that they are well clustered in the n-dimensional space where each spectrum is a point.  They out number in the linear mass regime of the scaling relations defined by Bernardi et al. (2010). Consequently, the scaling relations of \rm{EASK2} are expected to be representative of the "elliptical class". This is confirmed in Tab. \ref{tab3}, showing that the fitting parameters of the scaling relations of \rm{EASK2} and all ellipticals to be very similar, and consistent with previous published results. 

Concerning the SFH of these galaxies, Fig. \ref{fig:SFH} shows that 90$\%$ of the stars were already formed 10 Gyr ago.  This is the espected formation history of the stellar population of a typical monolithic elliptical galaxy (see, e.g., \cite{carretero2007}). We can conclude that our classification scheme  separates as EASK2 typical ellipticals  that follow the standard scaling relations and star formation histories.

In addition, the class of elliptical galaxies also contains by two slightly different populations: \rm{EASK0}, and \rm{EASK3} classes. This separation is a continuous transition (see Fig. \ref{mf}) since these two classes are not completely clustered in the parameter space as the \rm{EASK2}  (see Fig. 8 as in \cite{sanchezalmeida2010}). This does not mean that they are outliers of the main class. Galaxies from \rm{EASK0} and \rm{EASK3} indeed represent  $\sim$50$\%$ of all ellipticals. On average, \rm{EASK3} galaxies are bluer, larger, and less massive while \rm{EASK0} are slightly redder and more massive (see Figs. \ref{f1} and \ref{mf}). 

\subsection{\rm{EASK0}: dry major mergers origin?}

\rm{EASK0} galaxies become a significant fraction of the elliptical population at the high-mass end, where the curvature of the mass-size relation appears (\cite{bernardi2010}; see also Fig. \ref{f2}). In a more recent work, Bernardi et al. (2011)  argue that this curvature seen in the mass-size relation can be a consequence of major dry-mergers becoming dominant in the high-mass end of the mass distribution ($\log(M_{star})>11.3$). Since \rm{EASK0} galaxies are  more abundant at this mass, they are significantly contributing to the curvature. We  speculate that \rm{EASK0} galaxies could  be elliptical galaxies that have experienced more major dry merger events than the average population of ellipticals.

The mean SFH shows that a large fraction (about 90$\%$) of their stars were already in place 10 Gyr ago as for \rm{EASK2}. Nevertheless, they have slightly higher metallicity than \rm{EASK2} galaxies (which may also account for the redder colors seen in the color-mass diagram; Fig. \ref{f1}). This could imply that \rm{EASK0} galaxies formed stars in a more efficient way, in the sense that they consumed a larger fraction of the original gas leading to larger metallicities. Indeed, the difference of the SFH between \rm{EASK2} and \rm{EASK0} galaxies is positive, meaning that \rm{EASK0} formed their stars somewhat earlier (see Fig.\ref{fig:SFH}).

Based on the above arguments, we speculate that \rm{EASK0} galaxies could be representative of the brightest cluster galaxies (BCGs). These galaxies would be located in high galaxy density environments (mainly centers of galaxy clusters). In these special regions, dry mergers are normal as gas stripping efficiently removes the gas content of galaxies in short time-scales (see e.g., \cite{quilis2000}).  If the hypotheses out to be correct, our classification scheme is able to isolate, among the population of elliptical galaxies, those which have experienced more major dry mergers and formed stars more efficiently on average (see Fig.\ref{fspec}). Moreover, even if dominant at high masses (as pointed out by Bernardi et al. 2011), they are also present at all stellar masses. 

\begin{figure}
   \centering
   \includegraphics[width=9cm,height=7cm]{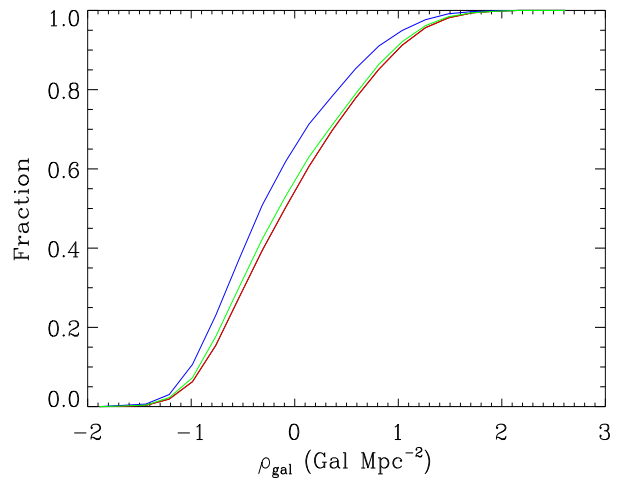}
   \caption{Cumulative distribution function of the galaxy density of the EASK0 (red), EASK2 (green), and EASK3 (blue) galaxies in the sample of Nair \& Abraham (2010).}
              \label{densidad}
    \end{figure}

We have investigated the environment of the EASK0, EASK2, and EASK3 galaxies in the sample of Nair \& Abraham (2010). Figure \ref{densidad} shows the galaxy density $(\rho_{gal})$ of the EASK classes. As explained in Nair \& Abraham (2010), the environmental overdensity was taken from Blanton et al. (2005). Thus, the galaxy neighbors of each galaxy of the catalog were considered as those galaxies in the magnitude range $M^{*} \pm 1$ and  within 5$h^{-1}$ Mpc excluding the target galaxy. Figure \ref{densidad} shows that EASK0 and EASK2 galaxies from
 Nair \& Abraham (2010) live in higher galaxy density environments than EASK3. We have run a Kolmogorov-Smirnov test, showing that the cumulative galaxy density distribution functions of EASK0 and EASK2 galaxies are statistically similar. In contrast, the cumulative galaxy density distribution function of the EASK3 galaxies is statistically different from those for EASK0 and EASK2.

\subsection{\rm{EASK3}: wet minor mergers?}

Galaxies in the color-mass diagram are thought to evolve from the blue cloud to the red sequence (see, e.g., \cite{kauffmann2006}). The bluer colors of \rm{EASK3} galaxies could possibly be caused by their recent arrival to  the red sequence.  If true, \rm{EASK3} should be poststarbust or E+A, galaxies. E+A galaxies show indeed a young stellar population together with a lack of on-going star formation. We have tested that possibility by determining the average value of the equivalent width (EW) of H$\delta$ and [OII] lines for  \rm{EASK0}, \rm{EASK2}, and \rm{EASK3}. This was done by using the EWs provided by MPA-JHU DR7 spectrum measurements. We obtain $<EW(H_{\delta})>=(-0.086\pm0.003) \rm{\AA}, (-0.049\pm0.002)\rm{\AA}$, and $(-0.0023\pm0.004) \rm{\AA}$ for \rm{EASK0}, \rm{EASK2}, and \rm{EASK3} galaxies respectively. On the other hand the $<EW([OII])>=(-2.99\pm0.03) \rm{\AA}, (-2.25\pm0.02) \rm{\AA},$ and $(-2.46\pm0.03) \rm{\AA}$ for \rm{EASK0}, \rm{EASK2}, and \rm{EASK3} galaxies respectively. These values rule out the hypothesis that \rm{EASK3} galaxies are dominated by E+A galaxies. This type of galaxies show $EW(H\delta)>5.0 \rm{\AA}$ and $EW([OII])> -2.5 \rm{\AA}$ (see \cite{goto2007}).

 \begin{figure}
   \centering
   \includegraphics[width=9cm,height=7cm]{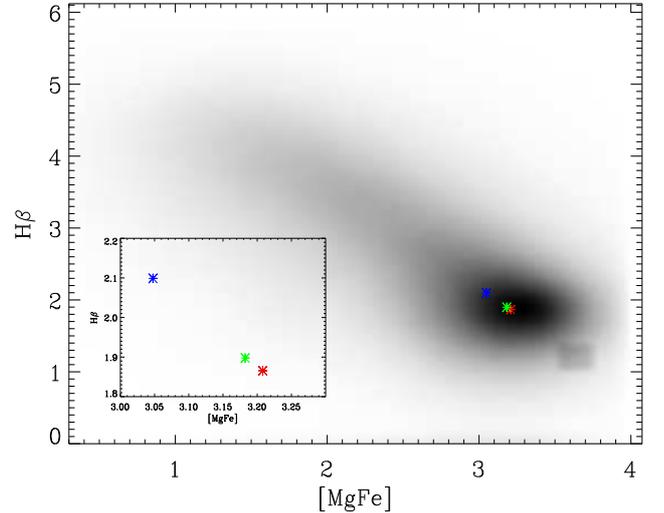}
   \caption{H$\beta$ versus [MgFe]$^{'}$ of the galaxies from SDSS-DR7 (grey scale). The asterisks show the mean values of the two indexes for \rm{EASK0} (red), \rm{EASK2} (green), and \rm{EASK3} (blue) galaxies.}
              \label{lickindex}%
    \end{figure}

Cold gas accretion into a galaxy halo can reach the central regions of the galaxy and turn on a central starburst (see \cite{dekel2009}). This is expected to happen at the intersection of filamentary structures of the Universe. Thus, this effect would be significant in galaxies with masses larger than $\approx 10^{12}$ M$_{\odot}$ (\cite{dekel2006}). Since \rm{EASK3} galaxies are essentially in the low mass end, we conclude that \rm{EASK3} galaxies are not likely rejuvened by direct gas accreation, and we are left with the possibility that they were formed by mergers of galaxies with  larger gas fraction than those forming \rm{EASK2} galaxies.

According to the structure formation scenario, mergers are the main driver in galaxy formation. In particular, the progenitors of elliptical galaxies could be both spheroidal or disk-like systems. The dissipative component of the progenitors plays an important role in determining the final properties of the remnants. Thus, during the merger process of gas-rich disk galaxies, part of the gas of the progenitors is driven towards the center of the remnant and feeds a central starburst that induces a rejuvenation of the central part of the final galaxy (see e.g. \cite{dimatteo2005}).  This new generation of central stars can  influence the structural parameters of the galaxies and produce changes in their scaling relations (see \cite{robertson2006}, \cite{hopkins2008}).  Therefore, we expect that elliptical galaxies built up after wet mergers, or those with a wet last merger, will show younger stellar populations in their centers than those formed by dry mergers. We have tested the possibility that galaxies from the \rm{EASK3} class are formed by mergers with a higher fraction of gas than those from \rm{EASK0} and \rm{EASK2} classes. In this case, we expect  \rm{EASK3} to have younger stellar populations in their central regions than  average. This seems to be happening according to the star formation histories represented in Fig. 11. In addition, we carried out an independent test by computing the mean age and metallicity of the galaxies from \rm{EASK0}, \rm{EASK2}, and \rm{EASK3} using two Lick spectral indexes: H$_{\beta}$ and [MgFe]$^{'}$. The H$_{\beta}$ index is an age-sensitive index, whereas [MgFe]$^{'}$ measures metalicity. Thomas et al. (2003) defined   [MgFe]$^{'}$ as

\begin{equation}
[\rm{MgFe}]^{'}=\sqrt{\rm{Mgb} \times (0.72 \rm{Fe5270} +0.28 \rm{Fe5335})},
\end{equation}

\noindent
where Mgb, Fe5270, and Fe5335 are Lick indexes.

[MgFe]$^{'}$ is a good tracer of the total metallicity of the stellar populations due to its independence of age and $\alpha$/Fe (see \cite{thomas2003}), and ensures to break down the age-metallicity degeneracy. The Lick indexes of the galaxies  were also obtained from MPA-JHU web page. The indexes used were those measured on the restframe spectrum after substraction of all 3$\sigma$ emission lines\footnote{see http://www.mpa-garching.mpg.de/SDSS/DR7}. Figure \ref{lickindex} shows the H$\beta$ versus [MgFe]$^{'}$ for the galaxies from SDSS-DR7 database. We have also overploted the mean values of these two indexes for EASK0, EASK2, and EASK3 galaxies. We have determined the stellar population properties (age and metallicity) of each elliptical galaxy class by using the program R-model (\cite{cardiel2003}). The result indicates that \rm{EASK3} galaxies are about 2 Gyr younger than galaxies from \rm{EASK0} and \rm{EASK2} classes. This is in full agreement with the SFH seen in Fig. \ref{fig:SFH} for these galaxies. It is evident from this figure that \rm{EASK3} galaxies have a more extended SFH than the canonical ellipticals (\rm{EASK2}). While only about 10$\%$ of the stars from \rm{EASK0} and \rm{EASK2} galaxies are formed in the last 8 Gyr, \rm{EASK3} galaxies formed up to 20$\%$.

Age and/or metallicity gradients have been observed in some early-type galaxies. These gradients, together with the aperture bias introduced by the 3\arcsec diameter of the SDSS fibers, could potentially create age differences between 
our classes of ellipticals. In order to discard this problem, we divided our galaxies in two groups: $r_{e} > 1\farcs 5$ and $r_{e} < 1\farcs 5$.
In both groups EASK3 galaxies are about 2 Gyr younger 
than EASK0 and EASK2, but the second group
is not sensitive to aperture bias since the 
SDSS fiber covers most of  galaxy.  Therefore, we discard
the difference of age being due to aperture bias.

Early-type galaxies can be produced by major mergers of equal-mass disks or spheroidal galaxies. Nevertheless, several minor mergers can also produce galaxies with photometrical and dynamical properties typical of early-type galaxies. In particular, repeated minor mergers can grow spheroidal galaxy components with Sersic shape parameter $n$ from  1 to 4 (\cite{aguerri2001}, \cite{eliche2006}, \cite{bournaud2007}) or produce galaxies with $V/\sigma$ similar to early-type galaxies (\cite{bournaud2007}). Minor mergers with large gas fraction can also activate central starbursts and produce the rejuvenation of the central regions of galaxies (\cite{bournaud2007}, \cite{kaviraj2009}). 

The structural parameters of the final remnant depend on the merger type, because the mass of the progenitors is deposited in a different way in major and minor mergers. Thus, major mergers can easily increase the mass density in the galaxy centers. In contrast, minor progenitors loose a large fraction of mass in the outer regions of the main progenitor (see \cite{aguerri2001}, \cite{eliche2006}). Robertson et al. (2006) and Hopkins et al. (2008) have investigated the scaling relations of early-type remnants formed by major disk-like galaxies with different gas fractions. They have shown that progenitors with a larger fraction of gas produce remnants following steeper size-mass relation and  showing larger central velocity dispersions. They also show that remnants of gas rich mergers are located in a FP with larger tilt than those from progenitors with small gas fractions. Our elliptical galaxies from \rm{EASK3} are located in a FP with larger tilt than \rm{EASK2} galaxies. This could be related to the larger gas fraction of the progenitors of the galaxies from \rm{EASK3}. In contrast, the size-mass and FJ relations of \rm{EASK3} galaxies show smaller slopes. This is not in agreement with the gas-rich major merger scenario modeled by Robertson et al. (2006) and Hopkins et al. (2008). We conclude that the rejuvenation of the central regions of the \rm{EASK3} galaxies does not seem to be caused by gas-rich major mergers. 

We can therefore conjeture that \rm{EASK3} galaxies have experienced more disipative minor mergers than the average ellipticals. Minor mergers increase indeed the effective radius and decrease the central velocity dispersion of the galaxies (see \cite{naab2009}). Our classification scheme can therefore provide the typical spectrum of this kind of galaxies (see Fig.\ref{fspec}). The disipative minor mergers undergone by EASK3 galaxies could also indicate that these galaxies are located in relatively less dense environments (see Fig. \ref{densidad}). This assumption is also in agreement with the percentages of the different classes of elliptical galaxies reported previously. Thus, it is well known that more than 50$\%$ of  galaxies are located in clusters or in group environments (see e.g., \cite{ramella2002}). This is particularly true for elliptical galaxies (see e.g., \cite{dressler1980}). This implies that more than 50$\%$ of elliptical galaxies are located in clusters or groups environments. We have seen in the previous sections that \rm{EASK0} and \rm{EASK2} galaxies are located in similar environments and represent about 80$\%$ of our ellipticals. In contrast, \rm{EASK3} galaxies are located in less dense environments are represent about 20$\%$ of the total sample. According with this numbers we can say that \rm{EASK3} galaxies could be located in the outskirts of galaxy clusters, or in field. In these environments gas stripping is not active and  small galaxies can retain their gas content. Thus, the environment could also explain the mass dichotomy observed in Fig. \ref{mf}. The role of environment in the age and metallicity of early-type galaxies was also observed by Clemens et al. (2006, 2009). Similarly to the results obtained here, they obtained that early-type galaxies located in field were younger than those located in clusters. They proposed that the assembling processes of elliptical galaxies are independent of the environment, although they are delayed in the field. This could be in agreement with our results if disipative minor mergers events are active until later times for EASK3 galaxies than for the other two classes.

\section{Conclusions}

We have isolated a sample of elliptical galaxies in the SDSS DR7 using an automated probabilty-based spectro-morphological classification.  We have shown that morphologically defined ellipticals are basically distributed in 3 spectral classes (\rm{EASK0}, \rm{EASK2}, \rm{EASK3}). For each of these three spectral classes, we have studied the classical scaling relations (color-mass, mass-size, Faber-Jackson, and fundamental-plane), as well as star formation histories, metallicities, and mass distributions.

The typical elliptical, understood as a galaxy which follows normal scaling relations of ellipticals, falls in our \rm{EASK2} class. This class represents $\sim 50\%$ of the whole population and essentially dominates at intermediate stellar masses. The bulk of their stellar content was already in place 10 Gyr ago. At the high mass end, \rm{EASK0} galaxies become more abundant. These ellipticals are slightly more metal rich and seem to have formed stars in a more efficient way than \rm{EASK2} galaxies. The tilt of the mass-size relation at high masses lead us to conclude that these galaxies could have experienced more dry mergers than the average. The low mass end is dominated by the \rm{EASK3} ellipticals. They are bluer, larger, and have smaller velocity dispersions at a fixed stellar mass. Moreover, they have a more extended stellar formation history, i.e., 20$\%$ of their stars were formed in the last 8 Gyr as compared to 10\% for the other classes. Minor gas rich mergers could be the main driver in the evolution of these objects if the rejuvenation of these galaxies comes from their central parts. This should be confirmed (at least fot the smallest galaxies) by forthcoming studies about stellar population gradients.

The environment could be at the base of the differences observed between the elliptical galaxy classes reported in this paper. EASK0 galaxies would be the brightest cluster galaxies, located in the centers of galaxy clusters were gas can be swept from galaxies in short time-scales due to gas stripping. Thus, dry galaxy mergers would be the main drivers of the evolution of galaxies located in these environments.  In contrast, \rm{EASK2} galaxies could be located in less dense environments where gas stripping mechanism is not active and galaxies can kept their gas content. 

Finally, we would like to emphasize that our classification can isolate the spectra of elliptical galaxies with different evolutive pathways, which could not be done with classical mass and/or color cuts. As a consequence, the present work can be used as a reference for studies at higher redshift. In forthcoming papers we will continue dissecting the spectro-morphological properties of local galaxies by extending this study to other galaxy classes.

\begin{acknowledgements}
      We would like to thank to J. Falcon Barroso for fruitful comments and discussion about stellar population in early-type galaxies. We are also in debt with R. Cid Fernandes for making available the SFH of the SDSS-DR7 galaxies. Thanks are due to the referee for helping us to improve both the contents and the presentation of the manuscript. This work was supported by the projects AYA2010-21887-C04-04 and by the Consolideer-Ingenio 2010 Program grant CSD2006-00070.    

 Funding for the Sloan Digital Sky Survey (SDSS) and SDSS-II has been provided by the Alfred P. Sloan Foundation, the Participating Institutions, the National Science Foundation, the U.S. Department of Energy, the National Aeronautics and Space Administration, the Japanese Monbukagakusho, and the Max Planck Society, and the Higher Education Funding Council for England. The SDSS Web site is http://www.sdss.org/.

    The SDSS is managed by the Astrophysical Research Consortium (ARC) for the Participating Institutions. The Participating Institutions are the American Museum of Natural History, Astrophysical Institute Potsdam, University of Basel, University of Cambridge, Case Western Reserve University, The University of Chicago, Drexel University, Fermilab, the Institute for Advanced Study, the Japan Participation Group, The Johns Hopkins University, the Joint Institute for Nuclear Astrophysics, the Kavli Institute for Particle Astrophysics and Cosmology, the Korean Scientist Group, the Chinese Academy of Sciences (LAMOST), Los Alamos National Laboratory, the Max-Planck-Institute for Astronomy (MPIA), the Max-Planck-Institute for Astrophysics (MPA), New Mexico State University, Ohio State University, University of Pittsburgh, University of Portsmouth, Princeton University, the United States Naval Observatory, and the University of Washington.

\end{acknowledgements}

\begin{appendix}
\section{Dependence of the results on the effective radius measurements}

We have estimated the effective radius of the galaxies as the radius containing 50\% of the total petrosian galaxy luminosity. This radius is not corrected for seeing, which may affect the scaling relations presented in this work. We have estimated this influence by measuring the mass-size relation of our E galaxies using the effective radius given by Simard et al. (2011). This different estimate was obtained by fitting the surface brightness distribution of the SDSS-DR7 galaxies with S\'esic models (see Simard et al. 2011). These models provide structural parameters of the galaxies after seeing correction. We have studied the influence of the used effective radius on the size-mass, fundamental plane, and Faber-Jackson relations, i.e., the relations involving $r_{e}$.

\begin{figure}
\centering
\includegraphics[width=9.0cm,height=7cm]{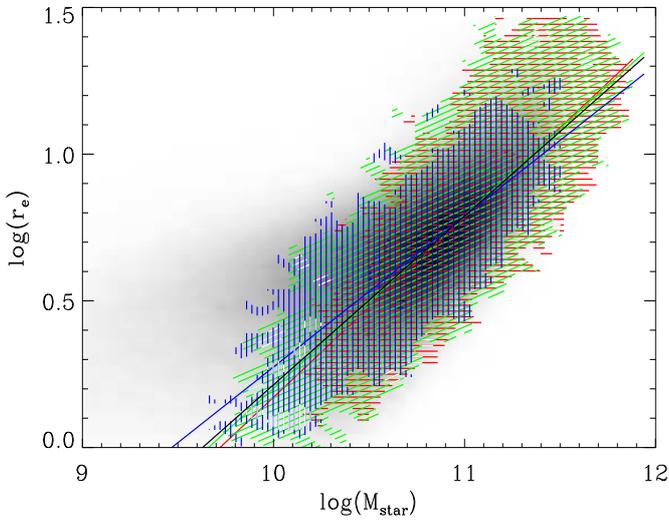}
\caption{Effective r-band radius from Simard et al. (2011) as a function of stellar mass for galaxies in the SDSS DR7 spectroscopic catalog (grey scale). The colors and lines are as in Fig. 7.
}
\label{appfig1}
\end{figure}

Figure \ref{appfig1} shows the mass-size relation for EASK0, EASK2, and EASK3 galaxies as computed in Sec. 3.2, but using the effective radius given by Simard et al. (2011). The fits of the size-mass relations showed in Fig. \ref{appfig1} are given in Tab. \ref{tabap}. Note that the relations obtained here are consistent with those obtained in Sec. 3.2. In particular, for a given stellar mass, galaxies from the EASK3 class are also larger. 

\begin{figure}
\centering
\includegraphics[width=9.0cm,height=7cm]{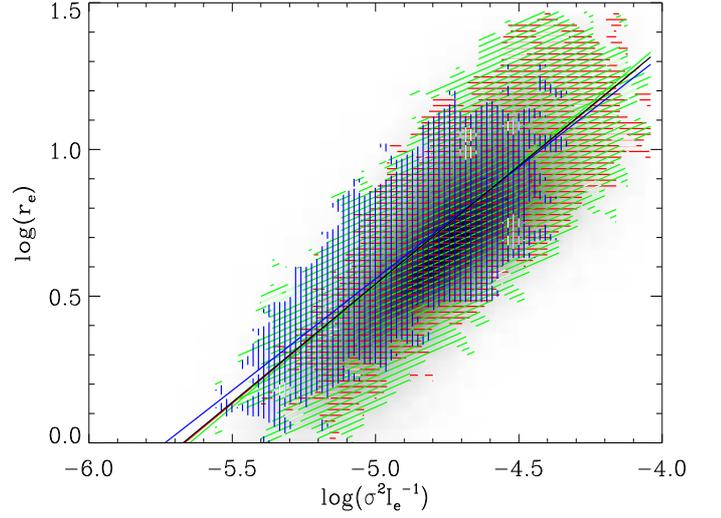}
\caption{Fundamental plane of the galaxies from SDSS-DR7 spectrosocpic catalog (grey color) using the effective radius given by Simard et al. (2011). 
The colors and lines are as in Fig. 8.
}
\label{appfig2}
\end{figure}

Figure \ref{appfig2} shows the fundamental plane (FP) for EASK0, EASK2, and EASK3 galaxies computed using the effective radius from Simard et al. (2011). The fits to these FPs are shown in Tab. \ref{tabap}. Notice that the behaviour is similar to that reported in Sec. 3.3. In particular, all classes present FPs with tilt, which is largest for EASK3 galaxies. In this case, the tilts of the FPs are closer to the virial plane ($b=1$) than those reported in Sec. 3.3.

\begin{figure}
\centering
\includegraphics[width=9.0cm,height=7cm]{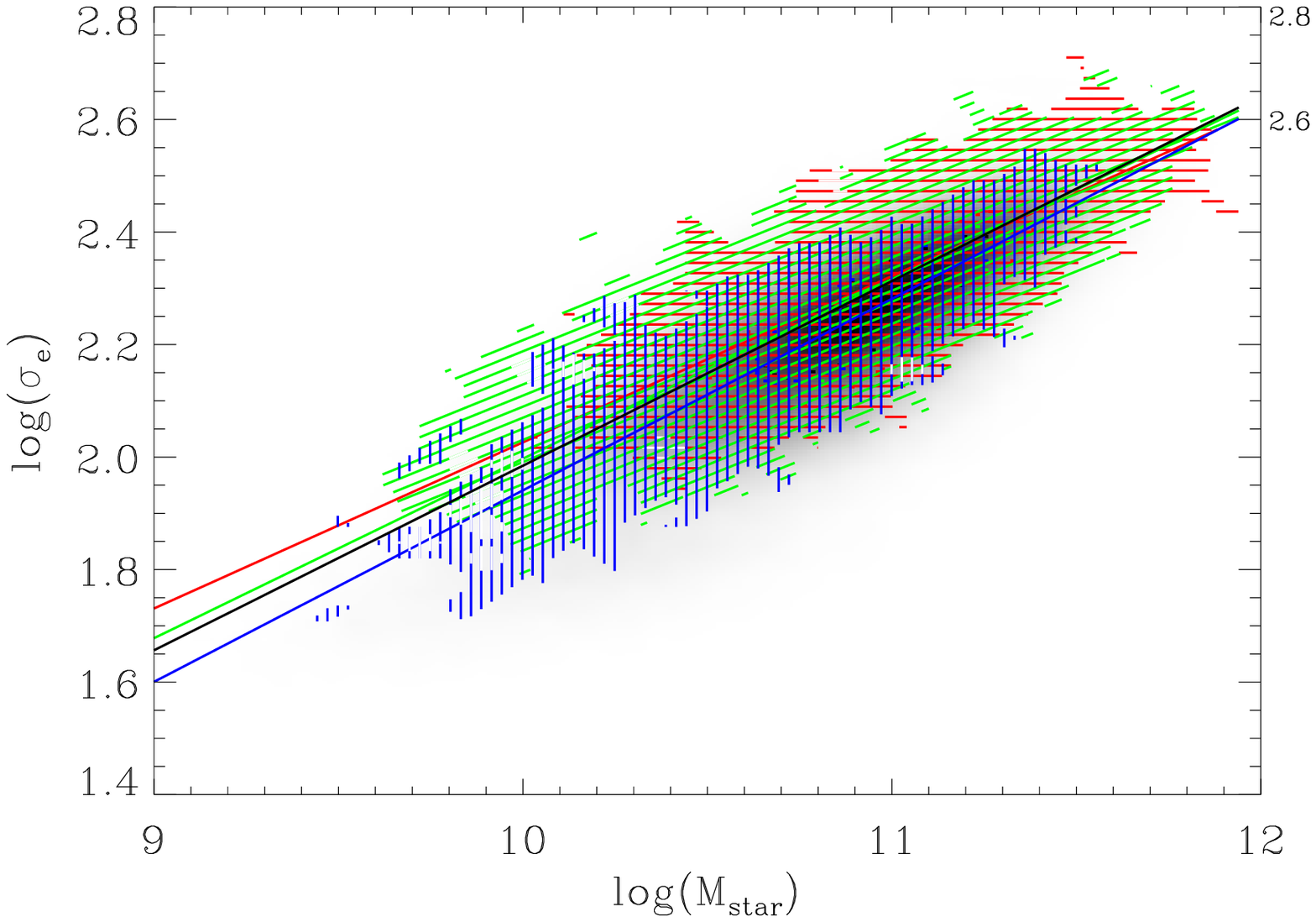}
\caption{The Faber-Jackson relation of the galaxies from SDSS-DR7 spectrosocpic catalog (grey color) using the effetive radius given by Simard et al. (2011). The colors and lines are as in Fig. 9.
}
\label{appfig3}
\end{figure}

Figure \ref{appfig3} and Tab. \ref{tabap} show the Faber-Jackson relations of EASK0, EASK2, and EASK3 galaxies derived using the effective radius of the galaxies from Simard et al. (2011). In this case, the behavior and fits to the relations are the same as those reported in Sec. 3.4. 

We can conclude that the results inferred from the scaling relationships  of EASK0, EASK2, and EASK3 do not depend on the way the effective radius is measured.

\begin{table}
\caption{Coefficients of linear fits to the size-mass, Faber-Jackson and FP relations for \rm{EASK0}, \rm{EASK2}, \rm{EASK3}, and as well as for all of them combined, using the effective  radii corrected for seeing by Simard et al. (2011).} 
\centering        
\begin{tabular}{c c c c} 
\hline\hline                 
Relation & Galaxy class & $a$ & $b$ \\  
\hline     
   Mass-size &\rm{EASK0} & -5.97$\pm$0.43 & 0.61$\pm$0.04  \\      
   & \rm{EASK2} & -5.70$\pm$0.26 & 0.59$\pm$0.02      \\
   & \rm{EASK3} & -4.87$\pm$0.43 & 0.51$\pm$0.04     \\
   & All & -5.54$\pm$0.13 & 0.57$\pm$0.01 \\
\hline
Faber-Jackson & \rm{EASK0} & -0.48$\pm$0.45& 0.25$\pm$0.04\\
& \rm{EASK2} & -0.72$\pm$0.28& 0.27$\pm$0.03\\
& \rm{EASK3} & -0.95$\pm$0.45&0.29$\pm$0.04 \\
& All & -0.81$\pm$0.14 & 0.28$\pm$0.01 \\
\hline
Fundamental Plane & \rm{EASK0} & 4.57$\pm$0.27 & 0.81$\pm$0.06 \\
& \rm{EASK2} & 4.68$\pm$0.18 & 0.82$\pm$0.04 \\
& \rm{EASK3} & 4.37$\pm$0.31 & 0.76$\pm$0.06 \\
& All & 4.57$\pm$0.09 & 0.81$\pm$0.01 \\ 
\hline                                   
\end{tabular}
\label{tabap}
\end{table}

\end{appendix}


\begin{thebibliography}{}

\bibitem[Abazajian et al. 2009]{abazajian2009} Abazajian, K.~N., et al.\ 2009, \apjs, 182, 543 

\bibitem[Abraham et al. 1996]{abraham1996} Abraham, R.~G., Tanvir,  N.~R., Santiago, B.~X., Ellis, R.~S., Glazebrook, K., \& van den Bergh, S.\ 1996, \mnras, 279, L47 

\bibitem[Aguerri et al. 2011]{aguerri2011} Aguerri, J.~A.~L., et al.\ 2011, \aap, 527, A143 

\bibitem[Aguerri \& Gonz{\'a}lez-Garc{\'{\i}}a 2009]{aguerri2009} Aguerri, J.~A.~L., \& Gonz{\'a}lez-Garc{\'{\i}}a, A.~C.\ 2009, \aap, 494, 891 


\bibitem[Aguerri et al. 2004]{aguerri2004} Aguerri, J.~A.~L., Iglesias-Paramo, J., Vilchez, J.~M., \& Mu{\~n}oz-Tu{\~n}{\'o}n, C.\ 2004, \aj, 127, 1344 

\bibitem[Aguerri \& Trujillo 2002]{Aguerri2002} Aguerri, J.~A.~L., \& Trujillo, I.\ 2002, \mnras, 333, 633 

\bibitem[Aguerri et al. 2001]{aguerri2001} Aguerri, J.~A.~L., Balcells, M., \& Peletier, R.~F.\ 2001, \aap, 367, 428 

\bibitem[Andreon \& Huertas-Company 2011]{andreon2011} Andreon, S., \& Huertas-Company, M.\ 2011, \aap, 526, A11 

\bibitem[Annibali et al. 2010]{annibali2010} Annibali, F., Bressan, A., Rampazzo, R., et al.\ 2010, \aap, 519, A40 

\bibitem[Asari et al. 2007]{asari2007} Asari, N.~V., Cid Fernandes, R., Stasi{\'n}ska, G., et al.\ 2007, \mnras, 381, 263 

\bibitem[Baldwin et al. 1981]{baldwin1981} Baldwin, J.~A., Phillips, M.~M., \& Terlevich, R.\ 1981, \pasp, 93, 5 

\bibitem[Balogh et al. 2004]{balogh2004} Balogh, M.~L., Baldry, 
I.~K., Nichol, R., Miller, C., Bower, R., 
\& Glazebrook, K.\ 2004, \apjl, 615, L101 

\bibitem[Barnes 1992]{barnes1992} Barnes, J.~E.\ 1992, \apj, 393, 484 

\bibitem[Barnes\& Hernquist 1991]{barnes1991} Barnes, J.~E., \& Hernquist, L.~E.\ 1991, \apjl, 370, L65 

\bibitem[Barnes \& Hernquist 1996]{barnes1996} Barnes, J.~E., \& Hernquist, L.\ 1996, \apj, 471, 115 


\bibitem[Bell et al. 2004]{bell2004} Bell, E.~F., et al.\ 2004, \apj, 608, 752 


\bibitem[Bell \& de Jong 2001]{bell2001} Bell, E.~F., \& de Jong, R.~S.\ 2001, \apj, 550, 212 

\bibitem[Bender et al. 1992]{bender1992} Bender, R., Burstein, D., \& Faber, S.~M.\ 1992, \apj, 399, 462 

\bibitem[Bernardi et al. 2011]{bernardi2011} Bernardi, M., Roche, 
N., Shankar, F., \& Sheth, R.~K.\ 2011, \mnras, 412, L6 

\bibitem[Bernardi et al. 2010]{bernardi2010} Bernardi, M., Shankar, F., Hyde, J.~B., Mei, S., Marulli, F., \& Sheth, R.~K.\ 2010, \mnras, 404, 2087 


\bibitem[Bernardi et al. 2003]{bernardi2003} Bernardi, M., et al.\ 2003, \aj, 125, 1817

\bibitem[Binette et al. 1994]{binette1994} Binette, L., Magris, C.~G., Stasi{\'n}ska, G., \& Bruzual, A.~G.\ 1994, \aap, 292, 13 



\bibitem[Blanton et al. 2005]{blanton2005} Blanton, M.~R., Eisenstein, D., Hogg, D.~W., Schlegel, D.~J., \& Brinkmann, J.\ 2005, \apj, 629, 143 


\bibitem[Bournaud et al. 2007]{bournaud2007} Bournaud, F., Jog, C.~J., \& Combes, F.\ 2007, \aap, 476, 1179 

\bibitem[Bressan et al. 2006]{bressan2006} Bressan, A., Panuzzo, 
P., Buson, L., et al.\ 2006, \apjl, 639, L55 

\bibitem[Bruzual \& Charlot 2003]{bruzual2003} Bruzual, G., \& Charlot, S.\ 2003, \mnras, 344, 1000 

\bibitem[Caon et al. 1993]{caon1993} Caon, N., Capaccioli, M., \& D'Onofrio, M.\ 1993, \mnras, 265, 1013 

\bibitem[Cappellari et al. 2007]{cappellari2007} Cappellari, M., et al.\ 2007, \mnras, 379, 418 

\bibitem[Cardelli et al. 1989]{cardelli1989} Cardelli, J.~A., Clayton, G.~C., \& Mathis, J.~S.\ 1989, \apj, 345, 245 

\bibitem[Cardiel et al. 2003]{cardiel2003} Cardiel, N., Gorgas, J., S{\'a}nchez-Bl{\'a}zquez, P., Cenarro, A.~J., Pedraz, S., Bruzual, G., \& Klement, J.\ 2003, \aap, 409, 511 


\bibitem[Carretero et al. 2007]{carretero2007} Carretero, C., Vazdekis, A., \& Beckman, J.~E.\ 2007, \mnras, 375, 1025 

\bibitem[Cid Fernandes et al. 2011]{cidfernandes2011} Cid Fernandes, 
R., Stasi{\'n}ska, G., Mateus, A., 
\& Vale Asari, N.\ 2011, \mnras, 413, 1687 

\bibitem[Cid Fernandes et al. 2007]{cidfernandes2007} Cid Fernandes, R., Asari, N.~V., Sodr{\'e}, L., et al.\ 2007, \mnras, 375, L16 


\bibitem[Cid Fernandes et al. 2005]{cidfernandes2005} Cid Fernandes, R., Mateus, A., Sodr{\'e}, L., Stasi{\'n}ska, G., \& Gomes, J.~M.\ 2005, \mnras, 358, 363 

\bibitem[Clemens et al. 2009]{clemens2009} Clemens, M.~S., 
Bressan, A., Nikolic, B., \& Rampazzo, R.\ 2009, \mnras, 392, L35 

\bibitem[Clemens et al. 2006]{clemens2006} Clemens, M.~S., 
Bressan, A., Nikolic, B., et al.\ 2006, \mnras, 370, 702 


\bibitem[Courteau et al. 2007]{courteau2007} Courteau, S., Dutton, A.~A., van den Bosch, F.~C., MacArthur, L.~A., Dekel, A., McIntosh, D.~H., \& Dale, D.~A.\ 2007, \apj, 671, 203 

\bibitem[Daddi et al. 2005]{daddi2005} Daddi, E., et al.\ 2005, 
\apj, 626, 680 

\bibitem[Dekel et al. 2009]{dekel2009} Dekel, A., Sari, R., \& Ceverino, D.\ 2009, \apj, 703, 785 

\bibitem[Dekel \& Birnboim 2006]{dekel2006} Dekel, A., \& Birnboim, Y.\ 2006, \mnras, 368, 2 


\bibitem[De Lucia et al. 2006]{delucia2006} De Lucia, G., Springel, V., White, S.~D.~M., Croton, D., \& Kauffmann, G.\ 2006, \mnras, 366, 499 


\bibitem[Desroches et al. 2007]{desroches2007} Desroches, L.-B., Quataert, E., Ma, C.-P., \& West, A.~A.\ 2007, \mnras, 377, 402 

\bibitem[Di Matteo et al. 2005]{dimatteo2005} Di Matteo, T., Springel, V., \& Hernquist, L.\ 2005, \nat, 433, 604 

\bibitem[Djorgovski \& Davis 1987]{djorgovsky1987} Djorgovski, S., \& Davis, M.\ 1987, \apj, 313, 59 


\bibitem[Dressler et al. 1987]{dressler1987} Dressler, A., Lynden-Bell, D., Burstein, D., Davies, R.~L., Faber, S.~M., Terlevich, R., \& Wegner, G.\ 1987, \apj, 313, 42 

\bibitem[Dressler 1980]{dressler1980} Dressler, A.\ 1980, \apj, 
236, 351 

\bibitem[Eliche-Moral et al. 2006]{eliche2006} Eliche-Moral, M.~C., Balcells, M., Aguerri, J.~A.~L., \& Gonz{\'a}lez-Garc{\'{\i}}a, A.~C.\ 2006, \aap, 457, 91 

\bibitem[Emsellem et al. 2007]{emsellen2007} Emsellem, E., et al.\ 2007, \mnras, 379, 401 

\bibitem[Faber et al. 1997]{faber1997} Faber, S.~M., et al.\ 1997, \aj, 114, 1771 


\bibitem[Faber \& Jackson 1976]{faber1976} Faber, S.~M., \& Jackson, R.~E.\ 1976, \apj, 204, 668 

\bibitem[Forbes et al. 1998]{forbes1998} Forbes, D.~A., Ponman, T.~J., \& Brown, R.~J.~N.\ 1998, \apjl, 508, L43 

\bibitem[Fritz et al. 2005]{fritz2005} Fritz, A., Ziegler, 
B.~L., Bower, R.~G., Smail, I., \& Davies, R.~L.\ 2005, \mnras, 358, 233 


\bibitem[Fukugita et al. 1998]{fukugita1998} Fukugita, M., Hogan,C.~ J., \& Peebles, P.~J.~E.\ 1998, \apj, 503, 518 

\bibitem[Gavazzi et al. 2005]{gavazzi2005} Gavazzi, G., Donati, A., Cucciati, O., Sabatini, S., Boselli, A., Davies, J., \& Zibetti, S.\ 2005, \aap, 430, 411 


\bibitem[Gerhard et al. 2001]{gerhard2001} Gerhard, O., 
Kronawitter, A., Saglia, R.~P., \& Bender, R.\ 2001, \aj, 121, 1936 

\bibitem[Girardi et al. 1996]{girardi1996} Girardi, L., Bressan, A., Chiosi, C., Bertelli, G., \& Nasi, E.\ 1996, \aaps, 117, 113 


\bibitem[Goto 2007]{goto2007} Goto, T.\ 2007, \mnras, 381, 187 

\bibitem[Graham\& Guzm{\'a}n 2003]{graham2003} Graham, A.~W., \& Guzm{\'a}n, R.\ 2003, \aj, 125, 2936 

\bibitem[Graves \& Faber 2010]{graves2010} Graves, G.~J., \& Faber, S.~M.\ 2010, \apj, 717, 803 

\bibitem[Graves et al. 2009]{graves2009} Graves, G.~J., Faber, S.~M., \& Schiavon, R.~P.\ 2009, \apj, 698, 1590 

\bibitem[Guti{\'e}rrez et al. 2004]{gutierrez2004} Guti{\'e}rrez, C.~M., Trujillo, I., Aguerri, J.~A.~L., Graham, A.~W., 
\& Caon, N.\ 2004, \apj, 602, 664 

\bibitem[Hernquist 1993]{hernquist1993} Hernquist, L.\ 1993, \apj, 409, 548 

\bibitem[Huertas-Company et al. 2011]{huertas2011} Huertas-Company, M., Aguerri, J.~A.~L., Bernardi, M., Mei, S., \& S{\'a}nchez Almeida, J.\ 2011, \aap, 525, A157 

\bibitem[Huertas-Company et al. 2010]{huertas2010} Huertas-Company, M., Aguerri, J.~A.~L., Tresse, L., Bolzonella, M., Koekemoer, A.~M., \& Maier, C.\ 2010, \aap, 515, A3 

\bibitem[Huertas-Company et al. 2008]{huertas2008} Huertas-Company, M., Rouan, D., Tasca, L., Soucail, G., \& Le F{\`e}vre, O.\ 2008, \aap, 478, 971 

\bibitem[Hopkins et al. 2009]{hopkins2009} Hopkins, P.~F., et al.\ 2009, \mnras, 397, 802 
\bibitem[Hopkins et al. 2008]{hopkins2008} Hopkins, P.~F., Cox, T.~J., \& Hernquist, L.\ 2008, \apj, 689, 17 

\bibitem[Hyde \& Bernardi 2009]{hyde2009} Hyde, J.~B., \& Bernardi, M.\ 2009, \mnras, 394, 1978 

\bibitem[Jerjen \& Binggeli 1997]{jerjen1997} Jerjen, H., \& Binggeli, B.\ 1997, The Nature of Elliptical Galaxies; 2nd Stromlo Symposium, 116, 239 

\bibitem[Jorgensen et al. 1995]{jorgensen1995} Jorgensen, I., Franx, M., \& Kjaergaard, P.\ 1995, \mnras, 276, 1341 

\bibitem[Kannappan et al. 2009]{kannappan2009} Kannappan, S.~J., Guie, J.~M., \& Baker, A.~J.\ 2009, \aj, 138, 579 

\bibitem[Kauffmann et al. 2006]{kauffmann2006} Kauffmann, G., 
Heckman, T.~M., De Lucia, G., Brinchmann, J., Charlot, S., Tremonti, C., 
White, S.~D.~M., \& Brinkmann, J.\ 2006, \mnras, 367, 1394 

\bibitem[Kauffmann et al. 2003]{kauffmann2003} Kauffmann, G., et 
al.\ 2003, \mnras, 341, 33 

\bibitem[Kaviraj et al. 2009]{kaviraj2009} Kaviraj, S., Peirani, S., Khochfar, S., Silk, J., \& Kay, S.\ 2009, \mnras, 394, 1713 


\bibitem[Kodama et al. 1998]{kodama1998} Kodama, T., Arimoto, N., Barger, A.~J., \& Arag'on-Salamanca, A.\ 1998, \aap, 334, 99 


\bibitem[Kodama \& Arimoto 1997]{kodama1997} Kodama, T., \& Arimoto, N.\ 1997, \aap, 320, 41 


\bibitem[Kormendy et al. 2009]{kormendy2009} Kormendy, J., Fisher, D.~B., Cornell, M.~E., \& Bender, R.\ 2009, \apjs, 182, 216 

\bibitem[Kormendy \& Bender 1996]{kormendy1996} Kormendy, J., \& Bender, R.\ 1996, \apjl, 464, L119 


\bibitem[Kormendy \& Illingworth 1982]{kormendy1982} Kormendy, J., \& Illingworth, G.\ 1982, \apj, 256, 460

\bibitem[Kormendy 1977]{kormendy1977} Kormendy, J.\ 1977, \apj, 218, 333

\bibitem[Krajnovi{\'c} et al. 2008]{krajnovic2008} Krajnovi{\'c}, D., et al.\ 2008, \mnras, 390, 93 


\bibitem[Kronawitter et al. 2000]{kronawitter2000} Kronawitter, A., Saglia, R.~P., Gerhard, O., \& Bender, R.\ 2000, \aaps, 144, 53

\bibitem[Lauer et al.2007]{lauer2007} Lauer, T.~R., et al.\ 2007, \apj, 664, 226 
 
\bibitem[Lauer et al. 2005]{lauer2005} Lauer, T.~R., et al.\ 2005, \aj, 129, 2138 

\bibitem[Larson 1975]{larson1975} Larson, R.~B.\ 1975, \mnras, 173, 671 

\bibitem[Lintott et al. 2011]{lintott2011} Lintott, C., et al.\ 
2011, \mnras, 410, 166 

\bibitem[Lintott et al. 2008]{lintott2008} Lintott, C.~J., et al.\ 
2008, \mnras, 389, 1179 

\bibitem[Masters et al. 2010]{masters2010} Masters, K.~L., et al.\ 2010, \mnras, 405, 783 

\bibitem[Mahajan \& Raychaudhury 2009]{mahajan2009} Mahajan, S., \& Raychaudhury, S.\ 2009, \mnras, 400, 687 


\bibitem[McIntosh et al. 2005]{mcintosh2005} McIntosh, D.~H., et al.\ 2005, \apj, 632, 191 

\bibitem[Mei et al. 2009]{mei09} Mei, S., et al.\ 2009, 
\apj, 690, 42 

\bibitem[M{\'e}ndez-Abreu et al. 2008]{mendez2008} M{\'e}ndez-Abreu, J., Aguerri, J.~A.~L., Corsini, E.~M., \& Simonneau, E.\ 2008, \aap, 487, 555 

\bibitem[Mihos \& Hernquist 1996]{mihos1996} Mihos, J.~C., \& Hernquist, L.\ 1996, \apj, 464, 641 

\bibitem[Morales-Luis et al. 2011]{moralesluis2011} Morales-Luis, 
A.~B., S{\'a}nchez Almeida, J., Aguerri, J.~A.~L., 
\& Mu{\~n}oz-Tu{\~n}{\'o}n, C.\ 2011, \apj, 743, 77 


\bibitem[Naab et al. 2009]{naab2009} Naab, T., Johansson, P.~H., \& Ostriker, J.~P.\ 2009, \apjl, 699, L178 


\bibitem[Naab et al. 2007]{naab2007} Naab, T., Johansson, P.~H., Ostriker, J.~P., \& Efstathiou, G.\ 2007, \apj, 658, 710 

\bibitem[Naab et al. 2006a]{naab2006a} Naab, T., Khochfar, S., \& Burkert, A.\ 2006a, \apjl, 636, L81 

\bibitem[Naab et al. 2006b]{naab2006b} Naab, T., Jesseit, R., \& Burkert, A.\ 2006b, \mnras, 372, 839 

\bibitem[Nair \& Abraham 2010]{nair2010} Nair, P.~B., \& Abraham, R.~G.\ 2010, \apjs, 186, 427 

\bibitem[Nieto et al. 1991]{nieto1991} Nieto, J.-L., Bender, R., Arnaud, J., \& Surma, P.\ 1991, \aap, 244, L25

\bibitem[Nigoche-Netro et al. 2010]{nigoche2010} Nigoche-Netro, A., Aguerri, J.~A.~L., Lagos, P., Ruelas-Mayorga, A., S{\'a}nchez, L.~J., \& Machado, A.\ 2010, \aap, 516, A96 

\bibitem[Nigoche-Netro et al. 2009]{nigoche2009} Nigoche-Netro, A., Ruelas-Mayorga, A., \& Franco-Balderas, A.\ 2009, \mnras, 392, 1060 

\bibitem[Nigoche-Netro et al. 2008]{nigoche2008} Nigoche-Netro, A., Ruelas-Mayorga, A., \& Franco-Balderas, A.\ 2008, \aap, 491, 731 

\bibitem[Padilla \& Strauss 2008]{padilla2008} Padilla, N.~D., \& Strauss, M.~A.\ 2008, \mnras, 388, 1321 

\bibitem[Partridge \& Peebles 1967]{partridge1967} Partridge, R.~B., \& Peebles, P.~J.~E.\ 1967, \apj, 147, 868 

\bibitem[Panuzzo et al. 2011]{panuzzo2011} Panuzzo, P., Rampazzo, R., Bressan, A., et al.\ 2011, \aap, 528, A10 

\bibitem[Persic et al. 1996]{persic1996} Persic, M., Salucci, P., \& Stel, F.\ 1996, \mnras, 281, 27 

\bibitem[Phillips et al. 1986]{phillips1986} Phillips, M.~M., Jenkins, C.~R., Dopita, M.~A., Sadler, E.~M., \& Binette, L.\ 1986, \aj, 91, 1062 






\bibitem[Quilis et al. 2000]{quilis2000} Quilis, V., Moore, B., \& Bower, R.\ 2000, Science, 288, 1617 

\bibitem[Ramella et al. 2002]{ramella2002} Ramella, M., Geller, 
M.~J., Pisani, A., \& da Costa, L.~N.\ 2002, \aj, 123, 2976 


\bibitem[Robertson et al. 2006]{robertson2006} Robertson, B., Cox, T.~J., Hernquist, L., Franx, M., Hopkins, P.~F., Martini, P., 
\& Springel, V.\ 2006, \apj, 641, 21

\bibitem[S{\'a}nchez-Almeida et al. 2011]{sanchezalmeida2011} S{\'a}nchez Almeida, J., Aguerri, J.~A.~L., Mu{\~n}oz-Tu{\~n}{\'o}n, C., \& Huertas-Company, M.\ 2011, \apj, in press

\bibitem[S{\'a}nchez Almeida et al. 2010]{sanchezalmeida2010} S{\'a}nchez Almeida, J., Aguerri, J.~A.~L., Mu{\~n}oz-Tu{\~n}{\'o}n, C., \& de Vicente, A.\ 2010, \apj, 714, 487 


\bibitem[Shen et al. 2003]{shen2003} Shen, S., Mo, H.~J., White, S.~D.~M., Blanton, M.~R., Kauffmann, G., Voges, W., Brinkmann, J., \& Csabai, I.\ 2003, \mnras, 343, 978

\bibitem[Simard et al. 2011]{simard2011} Simard, L., Mendel, J.~T., Patton, D.~R., Ellison, S.~L., \& McConnachie, A.~W.\ 2011, \apjs, 196, 11 

\bibitem[Stasi{\'n}ska et al. 2008]{stasinska2008} Stasi{\'n}ska, 
G., Vale Asari, N., Cid Fernandes, R., et al.\ 2008, \mnras, 391, L29 

\bibitem[Strateva et al. 2001]{strateva2001} Strateva, I., et al.\ 2001, \aj, 122, 1861


\bibitem[Terlevich \& Forbes 2002]{terlevich2002} Terlevich, A.~I., \& Forbes, D.~A.\ 2002, \mnras, 330, 547 


\bibitem[Thomas et al. 2003]{thomas2003} Thomas, D., Maraston, C., \& Bender, R.\ 2003, \mnras, 339, 897 

\bibitem[Thomas et al. 2007]{thomas2007} Thomas, J., Saglia, R.~P., Bender, R., Thomas, D., Gebhardt, K., Magorrian, J., Corsini, E.~M., \& Wegner, G.\ 2007, \mnras, 382, 657 

\bibitem[Toomre \& Toomre 1972]{toomre1972} Toomre, A., \& Toomre, J.\ 1972, \apj, 178, 623 


\bibitem[Tremblay \& Merritt 1995]{tremblay1995} Tremblay, B., \& Merritt, D.\ 1995, \aj, 110, 1039 

\bibitem[Trujillo \& Aguerri 2004]{trujillo2004} Trujillo, I., \& Aguerri, J.~A.~L.\ 2004, \mnras, 355, 82 

\bibitem[Trujillo et al. 2006]{trujillo2006} Trujillo, I., et al.\ 2006, \apj, 650, 18 

\bibitem[Trujillo et al. 2001]{trujillo2001} Trujillo, I., Aguerri, J.~A.~L., Guti{\'e}rrez, C.~M., \& Cepa, J.\ 2001, \aj, 122, 38 

\bibitem[van Dokkum \& Ellis 2003]{vandokum2003} van Dokkum, P.~G., \& Ellis, R.~S.\ 2003, \apjl, 592, L53 




\end{thebibliography}
\end{document}